\documentclass[aps,prb,twocolumn,superscriptaddress,eqsecnum,showpacs,showkeys,amsmath,amssymb]{revtex4}

\usepackage{amsfonts}
\usepackage{amssymb,amsmath}
\usepackage{graphicx}
\usepackage{dcolumn}
\usepackage{bm}
\usepackage{color}

\newcommand\trap{{\rm trap}}

\begin{document}

\title{Low-temperature properties of the Hubbard model
       on highly frustrated one-dimensional lattices}

\author{O. Derzhko}
\affiliation{Institute for Condensed Matter Physics,
          National Academy of Sciences of Ukraine,
          1 Svientsitskii Street, L'viv-11, 79011, Ukraine}
\affiliation{Max-Planck-Institut f\"{u}r Physik komplexer Systeme,
          N\"{o}thnitzer Stra\ss e 38, 01187 Dresden, Germany}
\affiliation{Institut f\"{u}r theoretische Physik,
          Universit\"{a}t Magdeburg,
          P.O. Box 4120, 39016 Magdeburg, Germany}
\author{J. Richter}
\affiliation{Institut f\"{u}r theoretische Physik,
          Universit\"{a}t Magdeburg,
          P.O. Box 4120, 39016 Magdeburg, Germany}
\author{A. Honecker}
\affiliation{Institut f\"{u}r Theoretische Physik,
          Georg-August-Universit\"{a}t G\"{o}ttingen,
          Friedrich-Hund-Platz 1, 37077 G\"{o}ttingen, Germany}
\author{M. Maksymenko}
\affiliation{Institute for Condensed Matter Physics,
          National Academy of Sciences of Ukraine,
          1 Svientsitskii Street, L'viv-11, 79011, Ukraine}
\author{R. Moessner}
\affiliation{Max-Planck-Institut f\"{u}r Physik komplexer Systeme,
          N\"{o}thnitzer Stra\ss e 38, 01187 Dresden, Germany}

\date{\today}



\begin{abstract}
We consider the repulsive Hubbard model on three highly frustrated
one-dimensional lattices -- sawtooth chain and two kagom\'{e} chains --
with completely dispersionless (flat) lowest single-electron bands. We
construct the complete manifold of {\em exact many-electron} ground
states at low electron fillings and calculate the degeneracy of these
states. As a result, we obtain closed-form expressions for
low-temperature thermodynamic quantities around a particular value of
the chemical potential $\mu_0$. We discuss specific features of
thermodynamic
quantities of these ground-state ensembles such as
residual entropy, an extra  low-temperature peak in the specific heat,
and the existence of ferromagnetism and paramagnetism.
We confirm our analytical
results by comparison with exact diagonalization data for finite systems.
\end{abstract}

\maketitle

\section{Introduction and motivation}
\label{sec1}
\setcounter{equation}{0}

The Hubbard model is a particularly simple model for strongly interacting electrons in
solids.\cite{thehubbard}
Nevertheless, rigorous analysis of the model is a difficult task
and exact statements about its properties
are notoriously rare.\cite{thehubbard,lieb,tasaki_jpcm}
A number of rigorous/exact results have been obtained for the Hubbard model
on some specific lattices,
see,
e.g.,
Refs.\ \onlinecite{mielke,tasaki,tasaki_ptp,tanaka_idogaki,batista,gulacsi,wu,dhr2007,dhr2009}.
In particular,
in the context of the origin of ferromagnetism in itinerant electron systems\cite{vollhardt}
different lattices supporting dispersionless (flat) single-electron band
were studied in some detail.\cite{mielke,tasaki,tasaki_ptp,tasaki_later,mielke_later}
In the last years  the theory of flat-band ferromagnetism
has been developed further.\cite{tanaka_ueda,sekizawa,tanaka_tasaki,watanabe,dots_wires,nishino,graphene}
Although at first glance one may think that the lattices admitting rigorous treatment are rather artificial,
nowadays new possibilities to design interacting lattice system with controlled geometry 
emerge.
Thus,
modern strategies in chemistry
open a route to synthesize new materials
with a desired lattice structure and intersite interaction.\cite{design}
Furthermore, recent progress in nanotechnology
allows the fabrication of quantum dot superlattices and quantum wire systems
with any type of lattice.\cite{dots_wires}
Another rapidly developing field
is the controlled setup of optical lattices for cold
atoms.\cite{cold_atom_a,cold_atoms,wu}

On the other hand,
during the past years it has been noticed
that exact ground states of the quantum $XXZ$ Heisenberg antiferromagnet
can be constructed at high magnetic fields
for a large class of geometrically frustrated lattices.\cite{localized_magnons}
These states,
called independent or isolated localized magnons,
are localized on nonoverlapping restricted areas of the lattice
and they clearly manifest themselves
in various peculiarities of the low-temperature strong-field properties of the spin systems
(macroscopic magnetization jump,\cite{localized_magnons}
field-tuned lattice instability,\cite{loc_mag_sp}
residual entropy,\cite{loc_mag_magnetocal,loc_mag_thermo1,loc_mag_thermo2}
enhanced magnetocaloric effect,\cite{loc_mag_magnetocal,schnack2007}
order-disorder phase transition below the saturation field\cite{loc_mag_thermo1,loc_mag_thermo2} etc.).
Interestingly,
flat-band ferromagnetism of Hubbard electrons exhibits some similarities
to the localized-magnon effect
for $XXZ$ Heisenberg antiferromagnets on certain frustrated lattices.
\cite{dhr2007,honecker_richter,dhr2009}
Note, however,
that while the one-particle description may be identical, the
many-particle picture is obviously different.
While the spin model  can be viewed as a hard-core bosonic system with nearest-neighbor intersite repulsion,
the electronic Hubbard model is a two-component fermionic system with on-site repulsion between different species.
Nevertheless, it has been found recently\cite{dhr2007,honecker_richter,dhr2009}
that several ideas developed for the Heisenberg model  can be carried over to the Hubbard model.

\begin{figure}
\begin{center}
\includegraphics[clip=on,width=\columnwidth,angle=0]{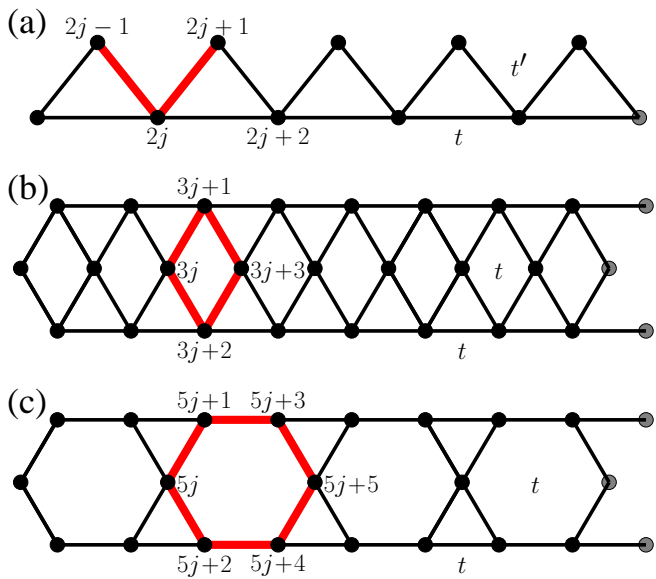}
\caption
{(Color online)
Three one-dimensional lattices considered in this paper:
(a) the sawtooth chain,
(b) the kagom\'{e} chain I,
and
(c) the kagom\'{e} chain II.
For the sawtooth Hubbard chain
the hopping parameter along the zig-zag path $t^{\prime}$
is $\sqrt{2}$ times larger than the hopping parameter $t>0$ along the base line.
For the kagom\'{e} Hubbard chains
all the hopping parameters $t>0$ are identical.
Bold (red) lines denote the minimal trapping cells for localized electrons.}
\label{fig01}
\end{center}
\end{figure}

In the present paper we consider the repulsive Hubbard model
on a class of one-dimensional frustrated lattices,
namely the sawtooth chain and two different  kagom\'{e} chains
(see Figs.~\ref{fig01}a, \ref{fig01}b, and \ref{fig01}c).
These Hubbard systems have highly degenerate ground states for certain
electron numbers $n\le n_{\max}$, $n_{\max} \propto N$,
where $N$ is the number of lattice sites.
We will give  explicit analytical expressions for all these ground states.
Their number grows exponentially with the system size
and can be counted 
by a mapping of the electronic problem onto a one-dimensional classical hard-dimer gas. 
Moreover,
these systems show saturated ground-state ferromagnetism for particular values of electron number
$n$, i.e., the square of the total spin is $\bm{S}^2=(n/2)(n/2+1)$.
Although the sawtooth chain on the one hand and the  kagom\'{e} chains
on the other hand belong to different types of flat-band
ferromagnets, the Hubbard model on all three lattices exhibits an
identical thermodynamic behavior
at low temperatures around a certain value of the chemical potential $\mu_0$
if $N\to\infty$.
Indeed, while the sawtooth lattice being a one-dimensional version of Tasaki's model
is an example of the ``cell construction'',\cite{tasaki,tasaki_ptp}
the kagom\'{e} chain I and the kagom\'{e} chain II belong to Mielke's class
of ``line
graphs'',\cite{mielke}
where the kagom\'{e} chain I is the line graph
of the two-leg ladder\cite{MSP03,KMTT09}
and the kagom\'{e} chain II is the line graph
of a decorated two-leg ladder.
The number of sites in the unit cell is 2, 3, and 5
for the sawtooth chain, kagom\'{e} chain I, and kagom\'{e} chain II,
respectively.
More differences can be seen in the single-electron energies for these lattices:
although the  tight-binding model of all three lattices exhibits a
dispersionless (flat) lowest-energy band, the next dispersive band is
separated by a finite gap for the sawtooth chain,
but it touches
the flat band at one point in momentum
space in the case of the kagom\'{e} chains
(for more details see Sec.~\ref{sec3}).
We mention that all of the three  lattices were discussed previously
in the context of various problems of strongly correlated systems,
see,
e.g.,
Refs.\ \onlinecite{sawtooth,pam,t-j,penc,sakamoto,waldtmann,martins,azaria,white,pati,indergand,chub2007}.

We consider the standard Hubbard Hamiltonian
\begin{eqnarray}
H&=&\sum_{\sigma=\uparrow,\downarrow}H_{0\sigma}
+H_{\rm{int}}
+\mu\sum_{i}\left(n_{i,\uparrow}+n_{i,\downarrow}\right),
\nonumber\\
H_{0\sigma}
&=&\sum_{\langle i,j \rangle}t_{i,j}
\left(c_{i,\sigma}^{\dagger}c_{j,\sigma}
+c_{j,\sigma}^{\dagger}c_{i,\sigma}\right),
\nonumber\\
H_{\rm{int}}&=&U\sum_{i}n_{i,\uparrow}n_{i,\downarrow}.
\label{1.01}
\end{eqnarray}
The sums run over $N$ lattice sites $i$ or over the nearest-neighbor pairs $\langle i,j\rangle$,
and periodic boundary conditions are imposed,
$c_{i,\sigma}^{\dagger}$ ($c_{i,\sigma}$)
are the usual fermion creation (annihilation) operators.
For the sawtooth chain
the hopping parameters along the zig-zag path $t^{\prime}$
are $\sqrt{2}$ times larger than the ones along the base line $t>0$
(then the lowest single-electron band is completely flat, see Sec.~\ref{sec3}).
For the kagom\'{e} chains
all the hopping parameters equal $t>0$
(then the lowest single-electron band is completely flat, see Sec.~\ref{sec3}).
$U\ge 0$ is the on-site Coulomb repulsion for electrons with different spins.
With further statistical-mechanics calculation in mind
we also introduce the term with chemical potential $\mu$.
Note that the sign of the term with $\mu$ in Eq.\ (\ref{1.01}) is chosen to have direct correspondence between
the chemical potential $\mu$ for the electronic model
and
the magnetic field $h$ for the respective antiferromagnetic spin
model.\cite{dhr2007,dhr2009,honecker_richter}
In what follows we often will set $t=1$.

The remainder of the paper is organized as follows.
We begin with a brief summary of our main results, Sec.~\ref{sec2}.
In Sec.~\ref{sec3} we discuss the tight-binding model of
non-interacting electrons [we set $U=0$ in Eq.\ (\ref{1.01})]
for the three chains.
In Sec.~\ref{sec4} we construct the complete set of exact many-electron ground states of the repulsive Hubbard model
on the three lattices
for the electron numbers $n\le n_{\max}$,
where $n_{\max}={\cal{N}}$ or ${\cal{N}}+1$ and ${\cal{N}}\propto N$,
and
discuss some properties of these states.
Moreover,
we explain the mapping of a certain subset of these states
onto spatial configurations of classical hard dimers on a simple chain.
This mapping is crucial to
calculate the exact degeneracies of the ground states for electron numbers
$n\le{\cal{N}}$
and to determine the residual entropy caused by these states.
In Sec.~\ref{sec5} we calculate analytically the contribution of the highly
degenerate  ground-state manifold
to the grand-canonical partition function of the Hubbard model on the
respective one-dimensional lattices.
This contribution dominates at low temperatures
when the chemical potential $\mu$ is around a certain value $\mu_0$
($\mu_0=2t$ for all three lattices).
We also calculate analytically the low-temperature behavior of several thermodynamic quantities
such as the average number of electrons, the entropy, and the specific heat,
and we
compare these analytical findings with numerical results obtained by exact diagonalization for finite
lattices.
Moreover, we use  exact diagonalization for finite
lattices
to discuss the influence of small deviations from the ideal geometry
(leading to a dispersion of the former flat band) on the low-temperature
thermodynamics.
In Sec.~\ref{sec6}
we use the analytical findings based on the  localized-state picture
and complementary numerical data from exact diagonalization of finite systems
for the discussion of the ground-state magnetic properties of the considered Hubbard chains.
In particular, we discuss the appearance of ferromagnetism and
paramagnetism.
In Sec.~\ref{xxz} we briefly discuss the relation between the electron
models and corresponding localized-spin models. 
Finally we summarize our findings in Sec.~\ref{sec7}.
Some auxiliary calculations are collected in Appendices.

\section{Summary of results}
\label{sec2}
\setcounter{equation}{0}

Compared to previous work on flat-band Heisenberg $XXZ$ magnets,
the theme of this paper is the new physics
which arises in itinerant magnets in which
(a) the mobile degrees of freedom are subject to the Pauli principle
and
(b) the Hamiltonian exhibits full SU(2) symmetry;
in particular, these features set this work apart from previous work on $XXZ$ models.

Previous studies on itinerant flat-band ferromagnets
were focused on particular values $n_{{\rm{f}}}$ of the electron number $n$ for which 
ground states with saturated ferromagnetism exist.
By contrast,
we characterize the complete set of ground states at electron numbers 
$n\le n_{{\rm{f}}}$, 
count their numbers, 
and use them to obtain explicitly the low-temperature thermodynamic quantities
as well as the average ground-state magnetic moments.

Our results cover, on an equal footing, two families of flat-band ferromagnets in one dimension.
The first are those obtained from Tasaki's cell construction 
and those described by Mielke's line-graph construction.

Somewhat unusually for a strongly interacting itinerant many-body system,
for these lattices we provide an explicit construction of the full set of exact ground states,
for a finite range of doping. The construction of the ground states is based
on a mapping to hard-core dimers on an appropriate one-dimensional structure.

This result enables us to obtain the corresponding partition functions,
and hence the low-temperature thermodynamics,
of the magnets in the low-doping regime.

The most salient consequences concern
(i) the entropy,
(ii) the low-temperature specific heat, 
(iii) the dependence of the average number of electrons on the chemical potential 
and
(iv) the magnetic properties.

Regarding (i) the flat band in the one-particle energies leads to
a huge degeneracy of the many-body ground states for a certain range of electron
densities resulting in a residual entropy in the thermodynamic limit.
This highly degenerate ground-state manifold has a great impact on the
low-temperature physics in the low-doping regime. In particular, an extra
low-temperature peak in the specific heat appears which is related to
an emerging low-energy scale separated from the energy scale 
determined by the value of the hopping integral.
Moreover, the zero-temperature average number of
electrons exhibits a jump at a certain value of the chemical potential.
Regarding (iv) 
the structure impressed on the many-body wave-function by the Pauli principle
leads to a degree of ferromagnetism in finite systems
which varies with electron number $n$;
there are specific fillings at which ferromagnetism is saturated
[the square of the total spin is $\bm{S}^2=(n/2)(n/2+1)$]
while at others it is only partially developed.
However, in the thermodynamic limit the region of electron density $n/N$, for
which ground-state ferromagnetism exists
shrinks to one point ($n/N=1/2$ for the sawtooth chain, $n/N=1/3$ for
the kagom\'{e} chain I, and $n/N=1/5$ for the  kagom\'{e} chain II).
For lower electron densities the ground state is paramagnetic and the
low-temperature behavior  
of the zero-field susceptibility follows a Curie law.

Finally we emphasize here
that the localized-magnon states for the $XXZ$ Heisenberg antiferromagnet
on all three lattices
can be also mapped onto a one-dimensional model of hard dimers.
Interestingly, 
due to the Pauli principle, the localized-electron states are
less constraining than the respective localized-magnon states. 
As a result the manifold of localized states for the
electronic system is
much larger than that for the magnon system.

\section{Non-interacting electrons. Trapped electron states}
\label{sec3}
\setcounter{equation}{0}

For zero Coulomb interaction $U$
the diagonalization of the Hamiltonian (\ref{1.01}) is straightforward.
Nevertheless, we start with the discussion of this case, since it provides
some important results which are relevant  for the case $U>0$, too.
The sawtooth chain has been studied in great detail before (see, e.g.,
Ref.~\onlinecite{honecker_richter}). We will therefore just recall the
main results for this case and give further details only for the two
kagom\'e chains.
For brevity, we may also omit spin indices as irrelevant in this section.

\subsection{Sawtooth chain}

\label{sec:IIIA}

The sawtooth chain consists of ${\cal{N}}=N/2$ cells,
each cell contains two sites,
see Fig.~\ref{fig01}a.
Hence,
there are two branches of single-particle energies
which read \cite{honecker_richter}
\begin{equation}
\label{3.01}
\varepsilon_{1,2}(\kappa)
=
t\cos\kappa
\mp\sqrt{t^2\cos^2\kappa+2{t^{\prime}}^2\left(1+\cos\kappa\right)}
+\mu. \quad
\end{equation}
For $t^{\prime}=\sqrt{2}t>0$
the lowest single-electron band becomes flat,
$\varepsilon_{1}(\kappa)=\varepsilon_{1}=-2t+\mu$ and one
can write down creation operators for a corresponding
set of eigenstates which are localized in a valley with
index $j$:\cite{honecker_richter}
\begin{equation}
\label{3.02}
l_{2j}^{\dagger} =
c_{2j-1}^{\dagger}-\sqrt{2}c_{2j}^{\dagger}+c_{2j+1}^{\dagger}\, .
\end{equation}
This set of localized single-electron states
is a convenient starting point
for the construction of the many-electron ground states of the Hamiltonian (\ref{1.01})
in the subspaces with electron numbers  $n=2,\ldots,{\cal{N}}$.
Moreover,
this ``localized'' point of view allows a useful simple geometrical interpretation,
see below.

\subsection{Kagom\'{e} chain I}

\label{sec:IIIB}

Consider next the kagom\'{e} chain I, see Fig.~\ref{fig01}b.
It consists of ${\cal{N}}=N/3$ cells,
each cell contains three sites.
After standard transformations we get the diagonal form of the Hamiltonian
\begin{eqnarray}
\label{3.03}
H_0+\mu \sum_i n_i
&=&\sum_{p=1}^3\sum_{\kappa}
\varepsilon_{p}(\kappa)\alpha^{\dagger}_{p,\kappa}\alpha_{p,\kappa},
\nonumber\\
\varepsilon_{1}(\kappa)
&=&
-2t+\mu,
\nonumber\\
\varepsilon_{2}(\kappa)
&=&
2t\cos\kappa+\mu,
\nonumber\\
\varepsilon_{3}(\kappa)
&=&
2t\left(1+\cos\kappa\right)+\mu
\end{eqnarray}
with $\kappa=2\pi m/{\cal{N}}$, $m \in \mathbb{Z}$,
$-{\cal{N}}/2 < m \le {\cal{N}}/2$.
The lowest energy band is flat,
$\varepsilon_{1}(\kappa)=\varepsilon_{1}=-2t+\mu$.
Note, however, that a state with $\kappa=\pi$
(it does exist if ${\cal{N}}$ is even)
from the dispersive band $\varepsilon_{2}(\kappa)$
has also the energy $\varepsilon_{1}=-2t+\mu$,
i.e., the next band touches the lowest flat band at $\kappa=\pi$.
The $\kappa$-dependent single-electron states
are given by $\alpha^{\dagger}_{p,\kappa}\vert 0\rangle$, $p=1,2,3$,
with
\begin{eqnarray}
\label{3.04}
\alpha^{\dagger}_{1,\kappa}
&=&-\frac{1}{\sqrt{2{\cal{N}}(2+\cos\kappa)}}\sum_{j=0}^{{\cal{N}}-1}
e^{-i\kappa j}l^{\dagger}_{3j},
\nonumber\\
l_{3j}^{\dagger}
&=&
c_{3j}^{\dagger}-c_{3j+1}^{\dagger}-c_{3j+2}^{\dagger}+c_{3j+3}^{\dagger},
\nonumber\\
\alpha^{\dagger}_{2,\kappa}
&=&
\frac{1}{\sqrt{2{\cal{N}}}}\sum_{j=0}^{{\cal{N}}-1}
e^{-i\kappa j} \left(c_{3j+1}^{\dagger}-c_{3j+2}^{\dagger}\right),
\nonumber\\
\alpha^{\dagger}_{3,\kappa}
&=&
\frac{1}{2\sqrt{{\cal{N}}(2+\cos\kappa)}}\sum_{j=0}^{{\cal{N}}-1}
e^{-i\kappa j}
\nonumber\\
&\times&
\left(c_{3j-2}^{\dagger}+c_{3j-1}^{\dagger}+2c_{3j}^{\dagger}+c_{3j+1}^{\dagger}+c_{3j+2}^{\dagger}\right).
\quad
\end{eqnarray}
Owing to the ${\cal{N}}$-fold degeneracy of the lowest flat band,
one can use alternatively the states
$l_{3j}^{\dagger}\vert 0\rangle$,
$j=0,1,\ldots,{\cal{N}}-1$
instead of the ${\cal{N}}$ states $\alpha^{\dagger}_{1,\kappa}\vert 0\rangle$.
These eigenstates are localized states where an electron is
trapped on a diamond consisting of four sites, $3j$, $3j+1$, $3j+3$, and
$3j+2$
($j$ enumerates these diamond traps and varies from 0 to ${\cal{N}}-1$).
It is easy to check that $[H_{0}+\mu\sum_{i}n_{i},l^{\dagger}_{3j}]=\varepsilon_{1}l^{\dagger}_{3j}$.
A real-space picture for the state with $\kappa=\pi$ from the dispersive band $\varepsilon_2(\kappa)$
is as follows:
\begin{eqnarray}
\label{3.05}
\alpha^{\dagger}_{2,\pi}\vert 0\rangle
=\frac{1}{\sqrt{2{\cal{N}}}}\sum_{j=0}^{{\cal{N}}-1}
(-1)^{j}
\left(c_{3j+1}^{\dagger}-c_{3j+2}^{\dagger}\right)\vert 0\rangle.
\end{eqnarray}
This eigenstate is not localized on a finite region,
but the electron is trapped on the two legs
[the sites inside the strip (i.e., numbers $3j$, $3j+3$, etc.)
do not appear in $\alpha^{\dagger}_{2,\pi}\vert 0\rangle$].
In what follows we call the state $\alpha^{\dagger}_{2,\pi}\vert 0\rangle$
the trapped two-leg state.
We may also introduce upper- and lower-leg states
\begin{eqnarray}
\label{3.06}
L^{\dagger}_u \vert 0\rangle
&=&\frac{1}{\sqrt{2}}\left(\alpha^{\dagger}_{2,\pi}-\lambda^{\dagger}\right)\vert 0\rangle,
\nonumber\\
L^{\dagger}_l \vert 0\rangle
&=&\frac{1}{\sqrt{2}}\left(\alpha^{\dagger}_{2,\pi}+\lambda^{\dagger}\right)\vert 0\rangle,
\nonumber\\
\lambda^{\dagger}&=&\frac{1}{\sqrt{2{\cal{N}}}}\sum_{j=0}^{{\cal{N}}-1}(-1)^{j}l_{3j}^{\dagger}.
\end{eqnarray}
Obviously,
the electron in the eigenstate $L^{\dagger}_u \vert 0\rangle$ ($L^{\dagger}_l \vert 0\rangle$)
is trapped on the upper (lower) leg.

\subsection{Kagom\'{e} chain II}

\label{sec:IIIC}

Finally,
we consider the kagom\'{e} chain II (Fig.~\ref{fig01}c)
which consists of ${\cal{N}}=N/5$ cells,
each cell contains five sites.
Again standard transformations lead to the diagonal form of the tight-binding  Hamiltonian
\begin{eqnarray}
\label{3.07}
H_0+\mu \sum_i n_i
&=&\sum_{p=1}^5\sum_{\kappa}
\varepsilon_{p}(\kappa)\alpha^{\dagger}_{p,\kappa}\alpha_{p,\kappa},
\nonumber\\
\varepsilon_{1}(\kappa)
&=&
-2t+\mu,
\nonumber\\
\varepsilon_{2}(\kappa)
&=&
- t\sqrt{2+2\cos\kappa}+\mu,
\nonumber\\
\varepsilon_{3}(\kappa)
&=&
t- t\sqrt{3+2\cos\kappa}+\mu,
\nonumber\\
\varepsilon_{4}(\kappa)
&=&
t\sqrt{2+2\cos\kappa}+\mu,
\nonumber\\
\varepsilon_{5}(\kappa)
&=&
t+ t\sqrt{3+2\cos\kappa}+\mu
\end{eqnarray}
with $\kappa=2\pi m/{\cal{N}}$, $m \in \mathbb{Z}$,
$-{\cal{N}}/2 < m \le {\cal{N}}/2$.
The state with the energy $\varepsilon_{2}(\kappa=0)$
(it does exist for odd and even ${\cal{N}}$)
touches the flat band.
The $\kappa$-dependent single-electron states
are given by
$\alpha^{\dagger}_{p,\kappa}\vert 0\rangle$, $p=1,2,3,4,5$.
For the sake of brevity
we give
$\alpha^{\dagger}_{1,\kappa}\vert 0\rangle$
for the lowest-energy band, only
\begin{eqnarray}
\label{3.08}
\alpha^{\dagger}_{1,\kappa}
=\frac{1}{\sqrt{2{\cal{N}}(3-\cos\kappa)}}\sum_{j=0}^{{\cal{N}}-1}
e^{-i\kappa j}l^{\dagger}_{5j},
\nonumber\\
l_{5j}^{\dagger}
=
c_{5j}^{\dagger}-c_{5j+1}^{\dagger}-c_{5j+2}^{\dagger}+c_{5j+3}^{\dagger}+c_{5j+4}^{\dagger}-c_{5j+5}^{\dagger}.
\end{eqnarray}
The corresponding localized states
are given by $l_{5j}^{\dagger}\vert 0\rangle$,
$j=0,1,\ldots,{\cal{N}}-1$, where an electron is
trapped on a hexagon
consisting of six sites, $5j, 5j+1, 5j+3, 5j+5, 5j+4, 5j+2$
($j$ enumerates these hexagon traps and varies from 0 to ${\cal{N}}-1$).
It is easily verified that $[H_{0}+\mu\sum_{i}n_{i},l^{\dagger}_{5j}]=\varepsilon_{1}l^{\dagger}_{5j}$.
A real-space picture for the state $\kappa=0$ from the dispersive band $\varepsilon_2(\kappa)$
is given by
\begin{equation}
\label{3.09}
\alpha^{\dagger}_{2,0}\vert 0\rangle
=\frac{1}{2\sqrt{{\cal{N}}}}\sum_{j=0}^{{\cal{N}}-1}
\left(c_{5j+1}^{\dagger}-c_{5j+2}^{\dagger}-c_{5j+3}^{\dagger}+c_{5j+4}^{\dagger}\right)\vert 0\rangle.
\end{equation}
As for the kagom\'{e} chain I
the electron in this eigenstate is trapped on the two legs.
Again in what follows we call the state $\alpha^{\dagger}_{2,0}\vert 0\rangle$
the trapped two-leg state.
Again we may introduce upper- and lower-leg states
\begin{eqnarray}
\label{3.10}
L^{\dagger}_u \vert 0\rangle
&=&\frac{1}{\sqrt{2}}\left(\alpha^{\dagger}_{2,0}-\lambda^{\dagger}\right)\vert 0\rangle,
\nonumber\\
L^{\dagger}_l \vert 0\rangle
&=&\frac{1}{\sqrt{2}}\left(\alpha^{\dagger}_{2,0}+\lambda^{\dagger}\right)\vert 0\rangle,
\nonumber\\
\lambda^{\dagger}&=&\frac{1}{2\sqrt{{\cal{N}}}}\sum_{j=0}^{{\cal{N}}-1}l_{5j}^{\dagger}.
\end{eqnarray}
Obviously,
the electron in the eigenstate $L^{\dagger}_u \vert 0\rangle$ ($L^{\dagger}_l \vert 0\rangle$)
is trapped on the upper (lower) leg.

\subsection{Trapped states and destructive interference. The geometrical perspective}

It is useful to discuss the appearance of localized electron states from a
geometrical  point of view.
As discussed above, for all three lattices we can easily single out a small
area of the lattice
which plays the role of a ``trapping cell'', namely
a {\sf V}-valley for the sawtooth chain, a diamond for the kagom\'{e} chain
I, and a hexagon for the kagom\'{e} chain II (see
marked regions in Fig.~\ref{fig01}).
Solving the single-electron problem for the trap
one finds the lowest-energy eigenfunction
$\propto\sum_ia_ic_i^{\dagger}\vert 0\rangle$
with
$a_1=1$, $a_2=-\sqrt{2}$, $a_3=1$
(sawtooth chain, the corresponding energy is $-\sqrt{2}t^{\prime}<0$),
$a_1=-a_2=a_3=-a_4=1$
(kagom\'{e} chain I, the corresponding energy is $-2t<0$),
or
$a_1=-a_2=a_3=-a_4=a_5=-a_6=1$
(kagom\'{e} chain II, the corresponding energy is $-2t<0$).
A crucial point is
that the scheme of the bonds connecting the trapping cell with the rest should prevent the escape of the localized
electron from the trap,
i.e., the constructed one-electron (localized) state
should remain an eigenstate of the Hamiltonian (\ref{1.01}) on the infinite lattice.
It is easy to show that a sufficient condition for this is
$\sum_{i}t_{r,i}a_i=0$,
where the sum runs over all sites $i$ of a trapping cell
and $r$ is an arbitrary site which does not belong to the trap,
see also Refs.~\onlinecite{localized_magnons} and \onlinecite{dhr2009}.
Indeed, the above condition is fulfilled
if an arbitrary bond belonging to the trap and the bonds attached to the two sites of this bond
in the trap
form such a triangle
that the electron amplitude on all sites outside the trapping cell is zero
(destructive quantum interference).
Note that a similar localization mechanism can be caused
by a magnetic field for tight-binding electrons in two-dimensional structures, 
where the wave packet is bounded in Aharonov-Bohm cages 
due to destructive interference for particular values of the magnetic flux.\cite{vidal}

Interestingly,
also the (extended) upper- and lower-leg states (\ref{3.06}), (\ref{3.10})
fit to this geometrical picture,
if we interpret each of the two legs as a regular polygon (with an even number of sites).
Then again two neighboring sites of the polygon are surrounded by equilateral triangles
which prevent the electron to escape from the leg.
Thus, for the kagom\'{e} chains
we have ${\cal{N}}$ localized states located on diamonds or hexagons
and, in addition, two  states trapped on the legs
(upper-leg state and lower-leg state),
i.e., in total ${\cal{N}}+2$ localized states.
(Note that for the sawtooth chain such additional states do not exist.)
However,
these ${\cal{N}}+2$ localized states are not linearly independent,
since there is one linear relation between them,
$L_l^{\dagger}\vert 0\rangle-L_u^{\dagger}\vert 0\rangle
=\sqrt{2}\lambda^{\dagger}\vert 0\rangle$,
where $L_l^{\dagger}$, $L_u^{\dagger}$, and $\lambda^{\dagger}$ are
given in Eq.\ (\ref{3.06}) or in Eq.\ (\ref{3.10})
(see also the general discussion of linear independence in Appendix~\ref{b}).
As a result,
there are only ${\cal{N}}+1$ linearly independent localized single-electron states.
These simple arguments are in perfect agreement
with the more detailed calculations presented in Sec.~\ref{sec:IIIB} and Sec.~\ref{sec:IIIC}.

\section{Trapped electron ground states for $U>0$}
\label{sec4}
\setcounter{equation}{0}

In the previous section,
we have found simple highly degenerate localized one-particle ground states
of noninteracting spinless electrons
which are created by operators
$l_{2j}^{\dagger}$, $l_{3j}^{\dagger}$, or $l_{5j}^{\dagger}$,
see Eqs.\ (\ref{3.02}), (\ref{3.04}), or (\ref{3.08})
[and also by $\alpha_{2,\pi}^{\dagger}$
or $\alpha_{2,0}^{\dagger}$ operators for the kagom\'{e} chains,
see Eqs.\ (\ref{3.05}) or (\ref{3.09})].
It is straightforward to create a set of $n$-electron ground states
with $1 <n \le {\cal{N}}$ for the sawtooth chain
and $1 <n \le {\cal{N}}+1$ for the kagom\'{e} chains
by applying $n$ different
(i.e., attached to different trapping cells)
operators
$l_{2j}^{\dagger}$, $l_{3j}^{\dagger}$, or $l_{5j}^{\dagger}$
[for the kagom\'{e} chain I (II) we may also apply the operator $\alpha^{\dagger}_{2,\pi}$ ($\alpha^{\dagger}_{2,0}$)].
The energy of these states is
$n\varepsilon_1$, and their degeneracy grows exponentially with the system size.

Now we return to the spinful case of interacting electrons,
i.e., $U>0$ in Eq.~(\ref{1.01}).
Clearly the inclusion of the spin  does not change the energy for $U=0$ but
increases the degeneracy. Let us denote the degeneracy at $U=0$ of the
ground states of $n$ electrons by $g^{(0)}_{{\cal{N}}}(n)$.
Obviously, one has $g^{(0)}_{{\cal{N}}}(n)={{2n_{\max}}\choose{n}}$,
$n\le n_{\max}$, where
$n_{\max}={\cal{N}}$ for the sawtooth chain and odd-${\cal{N}}$ kagom\'{e} chain I,
but
$n_{\max}={\cal{N}}+1$ for even-${\cal{N}}$ kagom\'{e} chain I and kagom\'{e} chain II.

First of all we note
that the Hubbard interaction in Eq.\ (\ref{1.01}) is a positive semidefinite operator
and hence it can only increase the eigenvalues
 of the Hamiltonian (\ref{1.01}).
On the other hand,
among the huge number of localized $n$-electron states being ground states for $U=0$
there is a considerable fraction of states
which do not feel the Hubbard interaction term
and thus they remain ground states with the $U$-independent energy $n\varepsilon_1$ for $U>0$.
However,
it is evident that the ground-state degeneracy, $g_{{\cal{N}}}(n)$, should decrease
as $U>0$ is switched on,
i.e., $g_{{\cal{N}}}(n)<g^{(0)}_{{\cal{N}}}(n)$.
We will consider the cases of the sawtooth chain and of the kagom\'{e} chains in more detail separately.

\subsection{Sawtooth chain and localized ground states}

\label{sec:IVA}

Let us recall that for a localized state
an electron with arbitrary spin $\sigma=\uparrow,\downarrow$ is trapped on three contiguous sites
({\sf V}-shaped trapping cell).
It is evident that $n$-electron states
where the electrons
(independently of their spins)
are located in disconnected {\sf V}-valleys
(i.e., {\sf V}-valleys without common sites)
are ground states in the $n$-electron subspace having the energy $n\varepsilon_1$.
The explicit expression for this type of ground states reads:
$l^{\dagger}_{2j_1,\sigma_1}\ldots l^{\dagger}_{2j_n,\sigma_n}\vert 0\rangle$,
$\forall \vert j_k-j_s \vert \ge 2$.

However, these states do not exhaust all ground states in the subspace with $n$ electrons.
Another type of ground states consists of $n$ trapped electrons
all with identical spin $\sigma$ occupying a cluster of
$n$ contiguous {\sf V}-valleys,
e.g.,
\begin{equation}
l^{\dagger}_{2j_1,\uparrow}\ldots l^{\dagger}_{2(j_1+n-1),\uparrow}\vert 0\rangle
\, .
\label{eq:1block}
\end{equation}
Since the interaction term is inactive, this is still an eigenstate of
the Hamiltonian (\ref{1.01}) with eigenvalue $n\varepsilon_1$.
Now we take into account the SU(2)-invariance of the Hubbard model (\ref{1.01}),
i.e.,
$\left[S^{-},H\right]=\left[S^{+},H\right]=\left[S^{z},H\right]=0$,
where
\begin{eqnarray}
S^{-} &=& \sum_{i}c_{i,\downarrow}^{\dagger}c_{i,\uparrow} \, , \nonumber \\
S^{+} &=& \sum_{i}c_{i,\uparrow}^{\dagger}c_{i,\downarrow} \, , \nonumber \\
S^{z} &=& \frac{1}{2}
\sum_{i}(c_{i,\uparrow}^{\dagger}c_{i,\uparrow}-c_{i,\downarrow}^{\dagger}c_{i,\downarrow})
\, .
\label{eq:defS}
\end{eqnarray}
Repeated application of the spin-lowering operator $S^{-}$ to the state
(\ref{eq:1block}) yields
$2s+1=n+1$ components of the spin-$(n/2)$ multiplet
\begin{equation}
\left(S^{-}\right)^m
l^{\dagger}_{2j_1,\uparrow}\ldots l^{\dagger}_{2(j_1+n-1),\uparrow}\vert 0\rangle,
\;\;\;
m=0,1,\ldots,n.
\label{eq:1blockS-}
\end{equation}
Evidently, all these states have the same energy $n\varepsilon_1$.
Using the commutation relation
\begin{eqnarray}
\label{4.01}
[S^{-},l_{2j,\uparrow}^{\dagger}]=l_{2j,\downarrow}^{\dagger},
\end{eqnarray}
one can easily write down the explicit expressions for
the states (\ref{eq:1blockS-})
in terms of operators $l_{2j}^{\dagger}$ only.

\begin{figure}
\begin{center}
\includegraphics[clip=on,width=\columnwidth,angle=0]{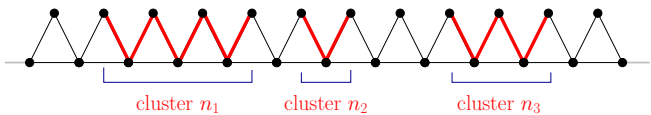}
\caption
{(Color online)
Illustration of a multi-cluster state. For each cluster
the total spin of the cluster can be flipped independently by applying the cluster
spin-flip operator
$S^{-}_{n_i}=\sum_{j \in {\rm cluster} \; n_i}
c_{j,\downarrow}^{\dagger}c_{j,\uparrow}$.}
\label{fig02}
\end{center}
\end{figure}

To construct the remaining ground states
in the subspace with $n$ electrons with the energy $n\varepsilon_1$ we
consider now multi-cluster states.\cite{dhr2007}
We consider all possible splittings of $n$ into a sum $n=n_1+n_2+\ldots;\,n_i>0$
such that clusters of $n_i$ consecutive occupied {\sf V}-valleys are
separated by at least one empty valley, see Fig.~\ref{fig02} for
an example.
Starting from a fully spin-polarized multi-cluster state
we can now independently act with the cluster spin-lowering operators,
$S^{-}_{n_i}=\sum_{j \in {\rm cluster} \; n_i}
c_{j,\downarrow}^{\dagger}c_{j,\uparrow}$,
where $j$ runs over all sites in the cluster $n_i$.
This yields many-electron ground states
which are products of the multiplet components in each cluster.

In order to compute the ground-state degeneracy $g_{{\cal{N}}}(n)$,
we now need to compute the number $D_{{\cal{N}}}(n)$ of states which
are constructed in the manner which we just described.
This is complicated a bit by the non-trivial components of the
SU(2)-multiplets consisting of linear combinations of products
of localized many-electron states. However, all coefficients
of these linear combinations are positive and one can choose
one state to represent the complete linear combination, for example, 
the one where all spins $\sigma=\uparrow$ are at the left
of each $n$ contiguously occupied {\sf V}-valleys and the
$\sigma=\downarrow$ are at the right of the clusters.\cite{dhr2007}
This reduces the counting problem to counting the number of
configurations ${\cal{Z}}(n,{\cal{N}})$
of three states in the trapping cells, namely empty ($0$) and
occupied with $\sigma=\uparrow,\downarrow$, subject to the
constraint that no $\uparrow$-state is allowed to appear
as the right neighbor of a $\downarrow$-state.
This combinatorial problem can be solved directly
with a $3\times3$ transfer matrix, yielding the
canonical partition functions ${\cal{Z}}(n,{\cal{N}})$
(see Appendix \ref{a}).
One small additional step then yields $D_{{\cal{N}}}(n)$:
in the sector with $n={\cal{N}}$ and for periodic boundary
conditions, there are only two allowed configurations according
to the rules which we just described whereas the SU(2)-multiplet
has ${\cal{N}}+1$ components (Tasaki's ferromagnetic ground states).
Hence, in the sector with $n={\cal{N}}$ and for periodic boundary
conditions we need to add ${\cal{N}}-1$ configurations by hand.
Putting all this together, we obtain the ground-state degeneracy
of the many-electron configurations with $n\le {\cal{N}}$ electrons localized in
${\cal{N}}$ traps
\begin{eqnarray}
\label{4.03}
g_{{\cal{N}}}(n)
&=&
D_{{\cal{N}}}(n) \, ,
\nonumber \\
D_{{\cal{N}}}(n) &=&
{\cal{Z}}(n,{\cal{N}})+({\cal{N}}-1)\,\delta_{n,{\cal{N}}}.
\end{eqnarray}
As described above, the last term, $({\cal{N}}-1)\,\delta_{n,{\cal{N}}}$,
ensures the correct counting in the sector $n=\cal N$.
Note furthermore that by writing $g_{{\cal{N}}}(n) = D_{{\cal{N}}}(n)$
we have implicitly assumed that the construction sketched at
the beginning of this sections yields all ground states.
This is indeed the case for the sawtooth chain, \cite{dhr2007}
see also below. 

An alternative way to compute ${\cal{Z}}(n,{\cal{N}})$ has been
described in Ref.~\onlinecite{dhr2007}: the combinatorial problem
of the three states $0$, $\sigma=\uparrow,\downarrow$
in ${\cal{N}}$ traps can
be mapped to a hard-dimer problem on an auxiliary simple chain with
$2{\cal{N}}$ sites. In this mapping, one associates
2 sites to each {\sf V}-valley in order to accommodate
either one spin projection such that the aforementioned constraint
that no $\uparrow$-state is allowed to appear
as the right neighbor of a $\downarrow$-state and
the constraint that double-occupancy of a {\sf V}-valley
is forbidden map to hard-dimer exclusion rules.
This auxiliary hard-dimer problem on $2{\cal{N}}$ sites
has been solved with a $2\times2$ transfer matrix\cite{dhr2007}
and therefore we may also refer to the trapped electron
configurations as ``hard-dimer configurations''. Of course,
the different ways of computing ${\cal{Z}}(n,{\cal{N}})$ are
completely equivalent.

In what follows, we call many-electron ground states,
which are constructed only from the localized single-electron states trapped
in a {\sf V}-valley ``hard-dimer states''.
Their degeneracy $D_{{\cal{N}}}(n)$ is given by
the second line of Eq.\ (\ref{4.03}).
The hard-dimer states are the only ground states
for the sawtooth chain and for the kagom\'{e} chain I with an odd number of cells ${\cal{N}}$.
For the kagom\'{e} chain I with an even number of cells ${\cal{N}}$ as well as for the kagom\'{e} chain II
in addition to the hard-dimer states
we have to consider also many-electron ground states
which involve an extended single-electron state with the flat-band energy $\varepsilon_1$
-- the two-leg state
[see Eqs.\ (\ref{3.05}) and (\ref{3.09})].

\subsection{Kagom\'{e} chains and trapped ground states involving two-leg states}

\label{sec:IVB}

Now we consider the ground states in the subspace with $n=1,2\ldots,{\cal{N}}+1$ electrons
for the case
when the two-leg state comes into play
(periodic kagom\'{e} chain I with even number of cells ${\cal{N}}$ or periodic
kagom\'{e} chain II).
Repeating the arguments elaborated for the sawtooth chain,
for $n=1,2,\ldots,{\cal{N}}$ we can construct hard-dimer states the number of which is given in Eq.\ (\ref{4.03}).
However, the hard-dimer states do not exhaust all many-electron ground states for $n=1,2,\ldots,{\cal{N}}$.
Indeed,
in the subspace with $n=1$ electron we have in addition two ground states
$\alpha^{\dagger}_{2,\pi,\sigma}\vert 0\rangle$, $\sigma=\uparrow,\downarrow$
for the kagom\'{e} chain I [see Eq.\ (\ref{3.05})]
or
$\alpha^{\dagger}_{2,0,\sigma}\vert 0\rangle$, $\sigma=\uparrow,\downarrow$
for the kagom\'{e} chain II [see Eq.\ (\ref{3.09})].

What happens if $n>1$?
Consider the case $n=2$.
We can construct ground states using the two-leg state as follows.
Assume both electrons have the same spin polarization
and the first one is localized within 1 of ${\cal{N}}$ (diamond or
hexagon) cells
whereas the second one is in a two-leg state.
Obviously this is the ground state with the energy $2\varepsilon_1$.
More states can be generated applying $S^{\pm}$ two times.
Thus, the degeneracy of the constructed ground state is $3{\cal{N}}$.

Interestingly,
in the subspace with $n=2$ electrons there is one more possibility
to construct an eigenstate of the interacting Hamiltonian (\ref{1.01})
with the energy $2\varepsilon_1$.
Recall that the electron in the state $L^{\dagger}_{u\sigma}\vert 0\rangle$
[see Eq.\ (\ref{3.06}) or Eq.\ (\ref{3.10})]
is located along the upper leg,
whereas the electron in the state $L^{\dagger}_{l\sigma}\vert 0\rangle$
[see Eq.\ (\ref{3.06}) or Eq.\ (\ref{3.10})]
is located along the lower leg.
Therefore, even if two electrons being in these states have opposite spins,
they do not feel the Hubbard repulsion (no common sites)
and the energy of this two-electron state remains $2\varepsilon_1$.
Thus, we find the following (extra) state in the subspace with $n=2$ electrons
for the periodic kagom\'{e} chains:
\begin{eqnarray}
\label{4.04}
\vert {\rm{extra}}\rangle
=L^{\dagger}_{l\sigma}L^{\dagger}_{u,-\sigma}
\vert 0\rangle,
\end{eqnarray}
where $L^{\dagger}_{u\sigma}$, $L^{\dagger}_{l\sigma}$
are defined by Eq.\ (\ref{3.06}) for the periodic kagom\'{e} chain I
and by Eq.\ (\ref{3.10}) for the periodic kagom\'{e} chain II.

Consider next the subspace with $n=3$ electrons.
Again we can assume three electrons to have the same spin,
two of which are localized within 2 of ${\cal{N}}$ cells
whereas the third one is in the two-leg state with the flat-band energy.
Obviously this is the ground state with the energy $3\varepsilon_1$.
More states can be generated applying $S^{\pm}$ three times.
Thus, the degeneracy of the constructed ground state is $4{{{\cal{N}}}\choose{2}}$.
Proceeding with such arguments for $n=4,\ldots,{\cal{N}}+1$ electrons
we can easily construct the ground states which involve the two-leg state.
Obviously, it is easy to count their number
which is equal to $(n+1){{\cal{N}}\choose{n-1}}$.

We wish to emphasize here
that for the kagom\'{e} chains
(kagom\'{e} I with even ${\cal{N}}$ or kagom\'{e} II chains)
we have constructed the ground state in the subspace with $n={\cal{N}}+1$ electrons,
which has the energy $({\cal{N}}+1)\varepsilon_{-}$,
starting from the fully polarized state
with ${\cal{N}}$ electrons occupying diamond or hexagon trapping cells,
respectively, and 1 electron being in the two-leg state,
and then applying the $S^{\pm}$ operator ${\cal{N}}+1$ times.
The degeneracy of the constructed ground state is ${\cal{N}}+2$.

Summing up,
we have arrived at the following formula for the degeneracy of the ground states
in the subspaces with $n\le {\cal{N}}+1$ electrons
for the periodic kagom\'{e} chain I with an even number of cells ${\cal{N}}$
or the periodic kagom\'{e} chain~II:
\begin{eqnarray}
\label{4.05}
g_{{\cal{N}}}(n)
&=&\left(1-\delta_{n,{\cal{N}}+1}\right)D_{\cal{N}}(n)+\left(1-\delta_{n,0}\right)L_{{\cal{N}}}(n),
\nonumber\\
L_{{\cal{N}}}(n)&=&(n+1){{\cal{N}}\choose{n-1}}+\delta_{n,2}
\end{eqnarray}
[$n=0,1,\ldots,{\cal{N}}$ for $D_{\cal{N}}(n)$,
whereas $n=1,\ldots,{\cal{N}}+1$ for $L_{\cal{N}}(n)$].

\subsection{Some properties of the ground states for $n\le n_{\max}$}

\label{sec:IVC}

As we have explained above,
the ground states in the subspaces with  $n\le n_{\max}$ electrons can be constructed
either from hard-dimer states only
(sawtooth, kagom\'{e} I with odd ${\cal{N}}$, $n_{\max}={\cal{N}}$)
or from hard-dimer states and the two-leg state
(kagom\'{e} I with even ${\cal{N}}$, kagom\'{e} II, $n_{\max}={\cal{N}}+1$)
and the
degeneracy of the ground state $g_{{\cal{N}}}(n)$ in the subspace with $n\le n_{\max}$ electrons
can be determined
either according to Eq.\ (\ref{4.03})
(sawtooth, kagom\'{e} I with odd ${\cal{N}}$)
or according to Eq.\ (\ref{4.05})
(kagom\'{e} I with even ${\cal{N}}$, kagom\'{e} II).
For the calculation of ${\cal{Z}}(n,{\cal{N}})$, see Appendix \ref{a}.

Fig.\ \ref{fig003} illustrates the difference in the ground-state degeneracy
$g_{{\cal{N}}}(n)$
conditioned by the ground states involving two-leg states.
We take as examples ${\cal{N}} =7$ and $8$.
The sawtooth chain never has two-leg state contributions,
the kagom\'{e} chain II always has two-leg state contributions,
and for the kagom\'{e} chain I 
the two-leg state contributions appear only for even ${\cal{N}}$.
Consequently, for ${\cal{N}}=8$, the degeneracies of the
two kagom\'{e} chains are identical whereas for
${\cal{N}}=7$ the degeneracy of the kagom\'{e} chain I is
identical to that of the sawtooth chain.
Firstly, we observe that the degeneracies increase rapidly
with ${\cal{N}}$ (note the logarithmic scale of the vertical axis
of Fig.\ \ref{fig003}). Secondly, there is an obvious contribution 
of the two-leg states.

\begin{figure}
\begin{center}
\includegraphics[clip=on,width=\columnwidth,angle=0]{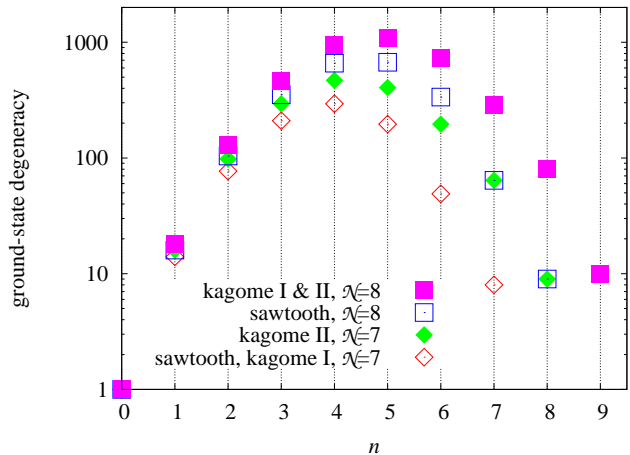}
\caption {(Color online)
Ground-state degeneracies for $n=1,\ldots,n_{\max}$
for
the sawtooth chain
(${\cal{N}}=7$, empty diamonds; ${\cal{N}}=8$, empty squares),
the kagom\'{e} chain I
(${\cal{N}}=7$, empty diamonds; ${\cal{N}}=8$, filled squares),
and
the kagom\'{e} chain II
(${\cal{N}}=7$, filled diamonds; ${\cal{N}}=8$, filled squares).}
\label{fig003}
\end{center}
\end{figure}
Next,
we discuss some general properties of the constructed many-electron ground states
in the subspaces with $n\le n_{\max}$ electrons.
As we have mentioned already,
these states having the energy $n\varepsilon_1$ as in the case $U=0$
are indeed the lowest-energy (ground) states for $U>0$
since the Hubbard interaction term $H_{\rm{int}}=U\sum_in_{i,\uparrow}n_{i,\downarrow}$ in Eq.\ (\ref{1.01})
is a positive semidefinite operator
and can only increase energies.

Another less simple question concerns completeness of the constructed ground states.
In other words,
are the constructed ground states the only ground states?
Here we use numerics for finite systems to check completeness.
Exact diagonalization data
for
the sawtooth chain with $N=6,\,8,\,10,\,12,\,14,\,16$,
the kagom\'{e} chain I with $N=9,\,12,\,15,\,18,\,21,\,24$,
and
the kagom\'{e} chain II with $N=15,\,20,\,25,\,30$
are in perfect agreement with predictions
of Sections \ref{sec:IVA} and \ref{sec:IVB},
compare also Fig.~\ref{fig003}.
These numerical findings are supported by general arguments:
We know that the localized states (including the two-leg states) are complete for $U=0$.
Adding a finite $U$, a part of these states will have higher energy,
but no new ground states will appear,
since $U$ can only increase the energy.

We have also estimated the energy gap between the ground state and the first excited state $\Delta_n$
in the subspace with $n$ electrons.
For $N\to\infty$ we have $\Delta_1=\varepsilon_2(\pi)-\varepsilon_1=2t$ (sawtooth chain) and $\Delta_1=0$ (kagom\'{e} chains).
Assuming $U \to \infty$ we found for finite systems of $N=30$ sites
$\Delta_2\approx 0.4044$ (sawtooth)
but
$\Delta_2\approx 0.0689$ (kagom\'{e} I)
and
$\Delta_2\approx 0.0584$ (kagom\'{e} II).
These data give a hint that the excitation gap for finite kagom\'{e} chains is much
smaller than for the sawtooth chain.
A finite-size extrapolation to $N \to \infty$ for fixed electron
density $n/N$
suggests a vanishing gap in the thermodynamic limit even for the sawtooth
chain.
Note, however, that in real materials with chain structure impurities and other lattice imperfections are always present 
and therefore one deals in practice with (an ensemble of) finite chains having a gap to the excited states.

With respect to the contribution of the trapped ground states 
(including the two-leg states) to thermodynamic quantities
the question arises whether
 these states for a given $n\le n_{\max}$ are linearly independent.
Their linear independence can be shown following the lines of Ref.\
\onlinecite{schmidt}, for more details see Appendix \ref{b}.

\section{Low-temperature thermodynamics}
\label{sec5}
\setcounter{equation}{0}

\subsection{Contribution of trapped electron states}

As shown in the preceding section,
for all three  one-dimensional lattices considered
the ground states of the Hubbard model (\ref{1.01}) in the subspaces with $n=0,1,\ldots,n_{\max}$ electrons
have the energy $n\varepsilon_1$ and the degeneracy $g_{{\cal{N}}}(n)$ [see Eqs.\ (\ref{4.03}), (\ref{4.05})].
Now if the chemical potential of electrons $\mu$ is around $\mu_0$ 
($\mu_0=2t$ for all three models)
the constructed ground states in the subspaces with $n=0,1,\ldots,n_{\max}$ electrons
will dominate the grand-canonical partition function at low temperatures due to their huge degeneracy,
i.e.,
\begin{eqnarray}
\Xi(T,\mu,N) &\approx& \Xi_{{\rm{GS}}}(T,\mu,N)
\nonumber\\
&=&\sum_{n=0}^{n_{\max}}g_{\cal{N}}(n)\exp\left(-\frac{n\varepsilon_1}{T}\right)
\nonumber\\
&=&\sum_{n=0}^{n_{\max}}g_{\cal{N}}(n)\exp\left[\frac{n(\mu_0-\mu)}{T}\right]
\nonumber\\
&=&\sum_{n=0}^{n_{\max}}g_{\cal{N}}(n)z^{n},
\label{5.01}
\end{eqnarray}
where $z=\exp x$ is the activity and $x=(\mu_0-\mu)/T$.

Consider first the sawtooth chain and the kagom\'{e} chain I with an odd number of 
trapping cells ${\cal{N}}$,
i.e., all ground states correspond to hard-dimer states.
Inserting Eq.\ (\ref{4.03}) into Eq.\ (\ref{5.01}) we arrive at
\begin{eqnarray}
\Xi_{{\rm{GS}}}(T,\mu,N)
&=&
\Xi_{{\trap}}(z,{\cal{N}})+({\cal{N}}-1)z^{{\cal{N}}},
\nonumber\\
\Xi_{{\trap}}(z,{\cal{N}})
&=&\sum_{n=0}^{{\cal{N}}}z^n{\cal{Z}}(n,{\cal{N}}).
\label{5.02}
\end{eqnarray}
Calculating $\Xi_{{\trap}}(z,{\cal{N}})$ with the help of the transfer-matrix method\cite{baxter}
(see Appendix \ref{a})
we get the following equations
\begin{eqnarray}
\Xi_{{\rm{GS}}}(T,\mu,N)
&=&
\xi_+^{\cal{N}}+\xi_-^{\cal{N}}+\xi_3^{\cal{N}},
\nonumber\\
\xi_{\pm}&=&\left(\frac{1}{2} \pm \sqrt{\frac{1}{4}+z}\right)^2,
\nonumber\\
\xi_3&=&({\cal{N}}-1)^{\frac{1}{{\cal{N}}}} z.
\label{5.03}
\end{eqnarray}
This is an important result, since it allows to calculate
the contribution of all the ground states described by hard dimers to
thermodynamic quantities explicitly.

Consider next the kagom\'{e} chain I with an
even number of trapping cells ${\cal{N}}$ and the kagom\'{e} chain II.
Insertion of Eq.\ (\ref{4.05}) into Eq.\ (\ref{5.01}) yields
\begin{eqnarray}
\Xi_{{\rm{GS}}}(T,\mu,N)
=
\Xi_{{\trap}}(z,{\cal{N}})+({\cal{N}}-1)z^{{\cal{N}}}
\nonumber\\
+\sum_{n=1}^{{\cal{N}}+1}(n+1){{\cal{N}}\choose{n-1}}z^n
+z^2.
\label{5.04}
\end{eqnarray}
Here the first two terms account for the many-electron configurations
associated to the ${\cal{N}}$ trapping cells,
cf.\ Eq.~(\ref{5.02}),
whereas the third and the fourth term are due to the two-leg state.
After simple calculations we get the following final result
\begin{eqnarray}
\Xi_{{\rm{GS}}}(T,\mu,N)
=
\xi_+^{\cal{N}}+\xi_-^{\cal{N}}+\xi_3^{\cal{N}}
+\xi_4^{\cal{N}}+\xi_5^{\cal{N}}+\xi_6^{\cal{N}},
\nonumber\\
\xi_4=(2z)^{\frac{1}{{\cal{N}}}}(1+z),
\;
\xi_5=({\cal{N}}z^2)^{\frac{1}{{\cal{N}}}}(1+z)^{\frac{{\cal{N}}-1}{{\cal{N}}}},
\;
\xi_6=z^{\frac{2}{{\cal{N}}}}
\nonumber\\
\label{5.05}
\end{eqnarray}
[$\xi_{\pm}$ and $\xi_3$ are defined in Eq.\ (\ref{5.03})].

The entropy $S(T,\mu,N)=-\partial \Omega(T,\mu,N)/\partial T$,
the grand-canonical specific heat
$C(T,\mu,N)\equiv T\partial S(T,\mu,N)/\partial T$,
and the average number of electrons $\overline{n}(T,\mu,N)=\partial \Omega(T,\mu,N)/\partial \mu$
follow from Eqs.\ (\ref{5.03}), (\ref{5.05}) 
and the formula for the grand-thermodynamical potential
$\Omega(T,\mu,N)=-T\ln\Xi(T,\mu,N)$.

Although explicit formulas for these thermodynamic quantities for finite systems
are too cumbersome to be written down explicitly here,
it might be useful to consider some limiting cases:\\
(i) For $T= 0$ and $\mu<\mu_0$
we have
$z\to \infty$ and consequently we find for the residual entropy 
$S(\mu,N)= \ln({\cal{N}}+1)$ and $\overline{n}(\mu,N)={\cal{N}}$
[hard-dimer states only, cf.\ Eq.~(\ref{5.03})]
or
$S(\mu,N)= \ln({\cal{N}}+2)$ and $\overline{n}(\mu,N)={\cal{N}}+1$ 
[hard-dimer states and two-leg states, cf.
Eq.(\ref{5.05})].\\
(ii) For $T = 0$ and $\mu>\mu_0$
we have
$z\to 0$ which leads to
$S(\mu,N)= 0$ and $\overline{n}(\mu,N)=0$.\\
(iii)
At $\mu=\mu_0$ we have $z= 1$ for all $T$
and
we find for the  residual entropy
$S(\mu_0,N)=\ln\{[(3+\sqrt{5})/2]^{{\cal{N}}}+[(3-\sqrt{5})/2]^{{\cal{N}}}+{\cal{N}}-1\}$
[hard-dimer states only, cf.\ Eq.~(\ref{5.03})]
or
$S(\mu_0,N)=\ln\{[(3+\sqrt{5})/2]^{{\cal{N}}}+[(3-\sqrt{5})/2]^{{\cal{N}}}+{\cal{N}}+2^{{\cal{N}}+1}+{\cal{N}}2^{{\cal{N}}-1}\}$
[hard-dimer states and two-leg states, cf. Eq.(\ref{5.05})]. 
Moreover,
$C(T,\mu_0,N)=0$ for any temperature independently of the system size.\\
Note, finally, that
the conventional (canonical) specific heat $C(T,n,N)$ for
a fixed number of electrons $n\le n_{\max}$
is identically zero within the localized-state description.

In the thermodynamic limit ${\cal{N}}\to \infty$
only the largest term, i.e., $\xi_{+}^{\cal N}$, survives
in the expressions (\ref{5.03}) and (\ref{5.05}) for the
partition function\cite{footnote_small} $\Xi_{{\rm{GS}}}(T,\mu,N)$ and we have
\begin{eqnarray}
\Xi_{{\rm{GS}}}(T,\mu,N)
=
\left(\frac{1}{2} + \sqrt{\frac{1}{4}+\exp
\frac{2t-\mu}{T}}\right)^{2\cal{N}},
\label{5.06}
\end{eqnarray}
which holds for all three considered one-dimensional lattices. The only
differences consist in the relation between the number of cells ${\cal{N}}$
and the lattice size $N$
(${\cal{N}}=N/2$, ${\cal{N}}=N/3$, and ${\cal{N}}=N/5$
for the sawtooth chain, the kagom\'{e} chain I, and the kagom\'{e} chain II,
respectively).
Explicit formulas for these thermodynamic quantities 
can be obtained easily from the simple expression (\ref{5.06}) for
$\Xi_{{\rm{GS}}}(T,\mu,N)$. 
For instance,  the residual entropy for ${\cal{N}}\to \infty$ is given by 
$S(\mu_0,N)/{\cal{N}}=\ln[(3+\sqrt{5})/2]=0.962\,42\ldots\;$, 
i.e., the ground-state degeneracy grows asymptotically with the system size
according to $\varphi^{2{\cal{N}}}$, 
where interestingly the golden mean $\varphi=(1+\sqrt{5})/2$ enters the expression.

\begin{figure}
\begin{center}
\includegraphics[clip=on,width=\columnwidth,angle=0]{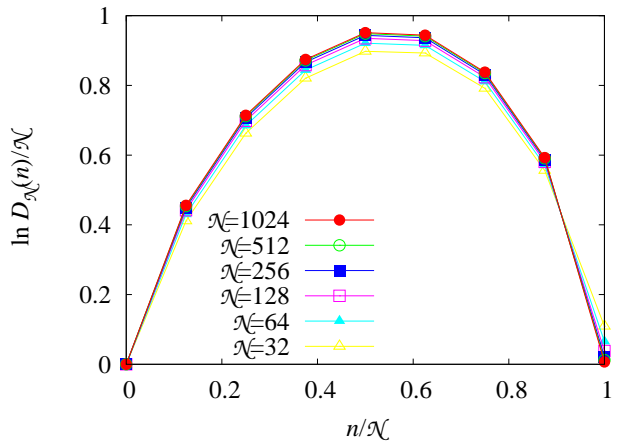}
\caption {(Color online)
Canonical residual entropy 
$S(n,{\cal{N}})/{\cal{N}}=\ln[D_{{\cal{N}}}(n)]/{\cal{N}}$ 
versus $n/{\cal{N}}$
for system sizes ${\cal{N}}=32,\,64,\,128,\,256,\,512,\,1024$.}
\label{fig03a}
\end{center}
\end{figure}

Using the transfer-matrix representation 
we can also calculate numerically the (canonical) residual entropy
$S(n,{\cal N})/{\cal N}=\ln[D_{{\cal{N}}}(n)]/{\cal N}$ related
to the hard-dimer states for certain fixed
values of the electron number $n$ for $\cal{N}$ up to 1024. 
The results are shown in Fig.~\ref{fig03a}.
From Fig.~\ref{fig03a} it is obvious that the largest degeneracy is
found for $n$ around ${\cal N}/2$. 
The extrapolation  
to $\cal{N} \to \infty$ yields 
$\lim_{\cal{N} \to
\infty}\ln[D_{{\cal{N}}}(n={\cal{N}}/2)]/{\cal{N}}\approx 0.955$, which is
very close to the grand-canonical residual entropy at $\mu=\mu_0$, see above.
Indeed, the maximum of $\lim_{\cal{N} \to\infty} S(n,{\cal N})/{\cal{N}}$ should
reproduce the grand-canonical residual entropy, albeit with slow finite-size
convergence.
 
\subsection{Comparison with exact diagonalization}

The above predictions are expected to be valid at low temperatures
and for $\mu$ around $\mu_0$. In order to verify these
expectations and examine the region of validity more precisely,
we have performed complementary exact diagonalization
of the full Hubbard Hamiltonian (\ref{1.01}). First,
we perform a symmetry reduction of the problem.
In particular, we use separate conservation of the
number of electrons with a given spin projection
$n_\uparrow$ and $n_\downarrow$ as well as translational
invariance. Ground states can then
be obtained with the help of the Lanczos recursion.\cite{Lanczos1950,CuWi}
In order to obtain thermodynamic properties, we perform a
full diagonalization in each symmetry subspace using a
library routine. The problem simplifies a bit in the
limit $U \to \infty$ where states with doubly occupied sites
can be eliminated from the Hilbert space. Nevertheless,
a full diagonalization of the complete problem can only
be performed for small $N$. Low-temperature approximations
for somewhat bigger $N$ can be obtained by omitting certain
sectors of $n$.

\begin{figure}
\begin{center}
\includegraphics[clip=on,width=\columnwidth,angle=0]{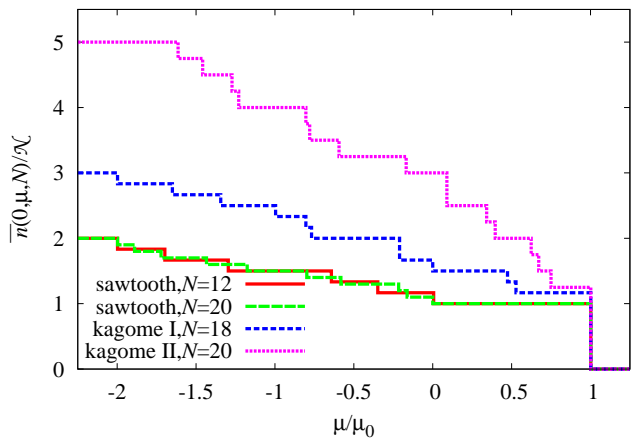}
\caption
{(Color online)
Average number of electrons in the ground state $\overline{n}(0,\mu,N)/{\cal{N}}$
versus
chemical potential $\mu/\mu_0$
for
the sawtooth chain [$N=12$ (solid), $N=20$ (long-dashed)],
the kagom\'{e} chain I [$N=18$ (short-dashed)],
the kagom\'{e} chain II [$N=20$ (dotted)]
with $U\to\infty$.
The result for $\overline{n}(0,\mu,N)/{\cal{N}}$ which follows from Eqs.\ (\ref{5.03}) and (\ref{5.05})
is given by $\theta(1-\mu/\mu_0)$ and $[({\cal{N}}+1)/{\cal{N}}]\theta(1-\mu/\mu_0)$, respectively.
}
\label{fig03}
\end{center}
\end{figure}

Fig.~\ref{fig03} shows exact diagonalization data
for the average number of electrons per cell $\overline{n}(T,\mu,N)/{\cal{N}}$ at $T=0$
as a function of the chemical potential $\mu/\mu_0$
for
the sawtooth chain ($N=12,\,20$),
the kagom\'{e} chain I ($N=18$),
and
the kagom\'{e} chain II ($N=20$).

\begin{figure}
\begin{center}
\includegraphics[clip=on,width=\columnwidth,angle=0]{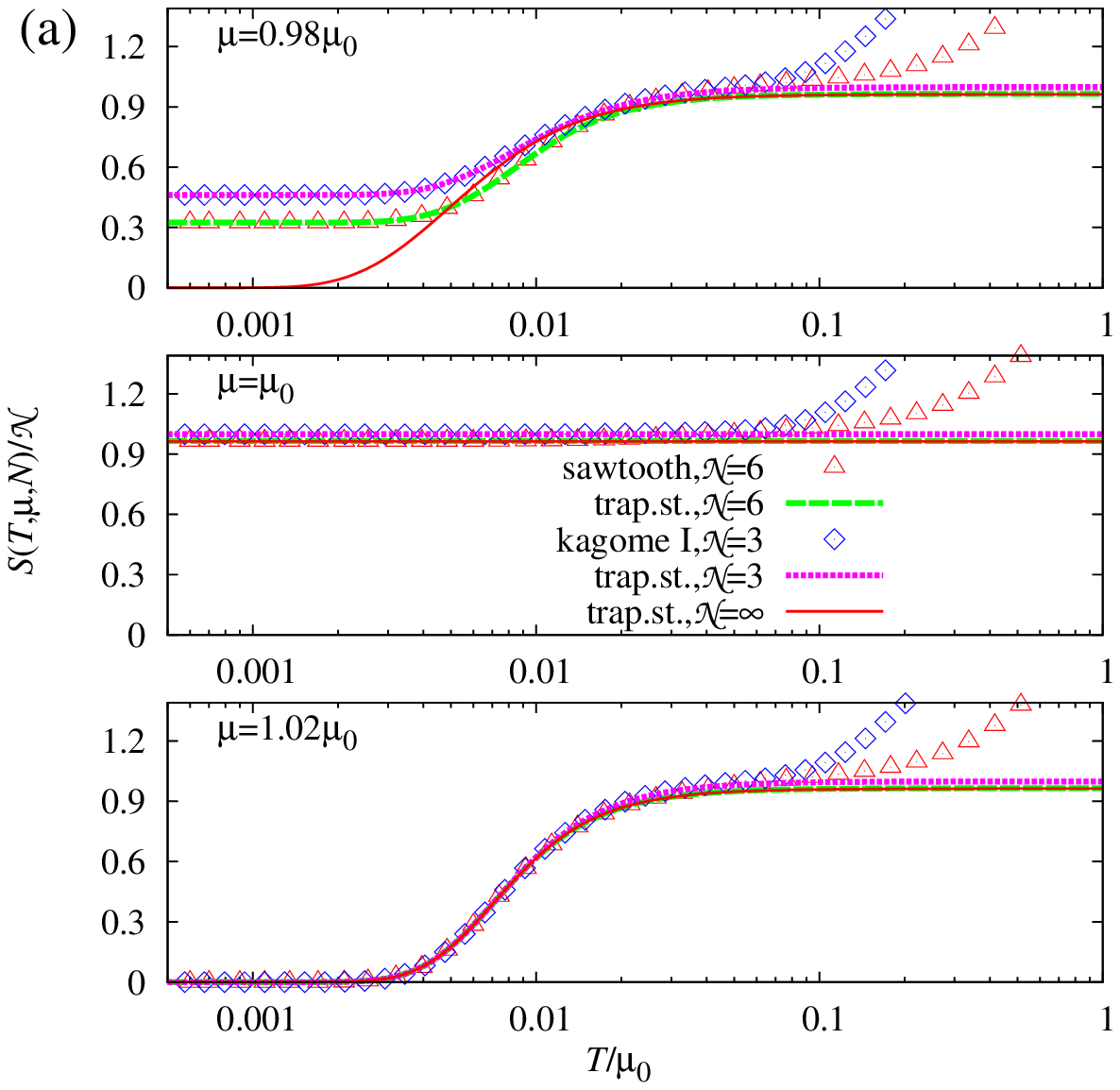}\\
\includegraphics[clip=on,width=\columnwidth,angle=0]{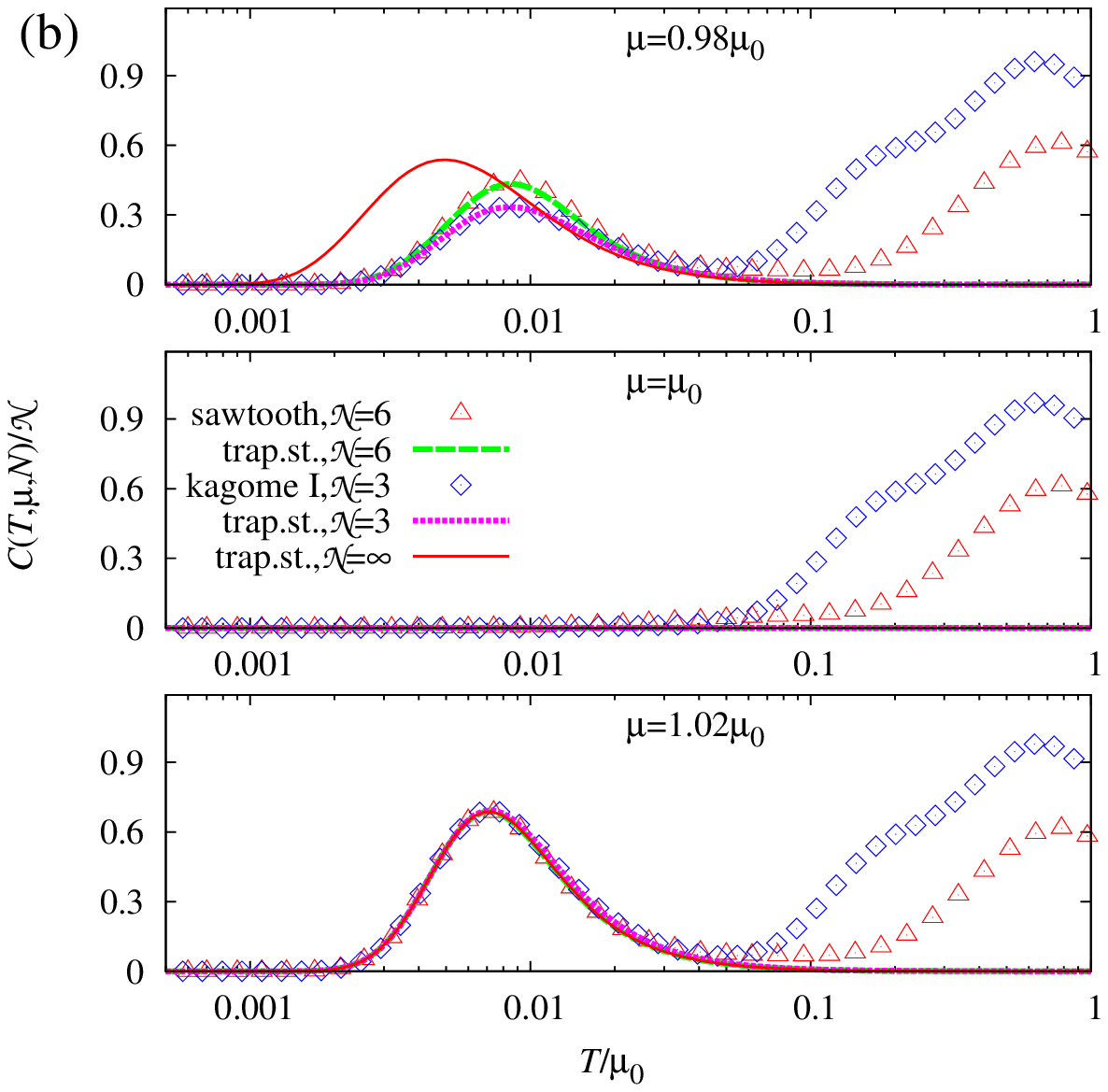}
\caption
{(Color online)
Entropy $S(T,\mu,N)/{\cal{N}}$
and
grand-canonical specific heat $C(T,\mu,N)/{\cal{N}}$
for
the sawtooth chain ($N=12$, triangles)
and
the kagom\'{e} chains I ($N=9$, diamonds)
with $t=1$, $U\to\infty$.
(a)
$S(T,\mu,N)/{\cal{N}}$ versus temperature $T/\mu_0$ at $\mu=0.98\mu_0,\;\mu_0,\;1.02\mu_0$ (from top to bottom).
(b)
$C(T,\mu,N)/{\cal{N}}$ versus temperature $T/\mu_0$ at $\mu=0.98\mu_0,\;\mu_0,\;1.02\mu_0$ (from top to bottom).
We also show
the hard-dimer result for $N\to\infty$ as it follows from (\ref{5.06})
(thin solid lines)
as well as
the results which follow from Eq.\ (\ref{5.03})
for ${\cal{N}}=3$ (dotted lines) and ${\cal{N}}=6$ (dashed lines).
Note that for these systems no additional leg states exist. 
Note further that
for $\mu=\mu_0$ and $\mu=1.02\mu_0$ the hard-dimer results for ${\cal N}=3$,
$6$, and $\infty$ are indistinguishable.}
\label{fig04}
\end{center}
\end{figure}

\begin{figure}
\begin{center}
\includegraphics[clip=on,width=\columnwidth,angle=0]{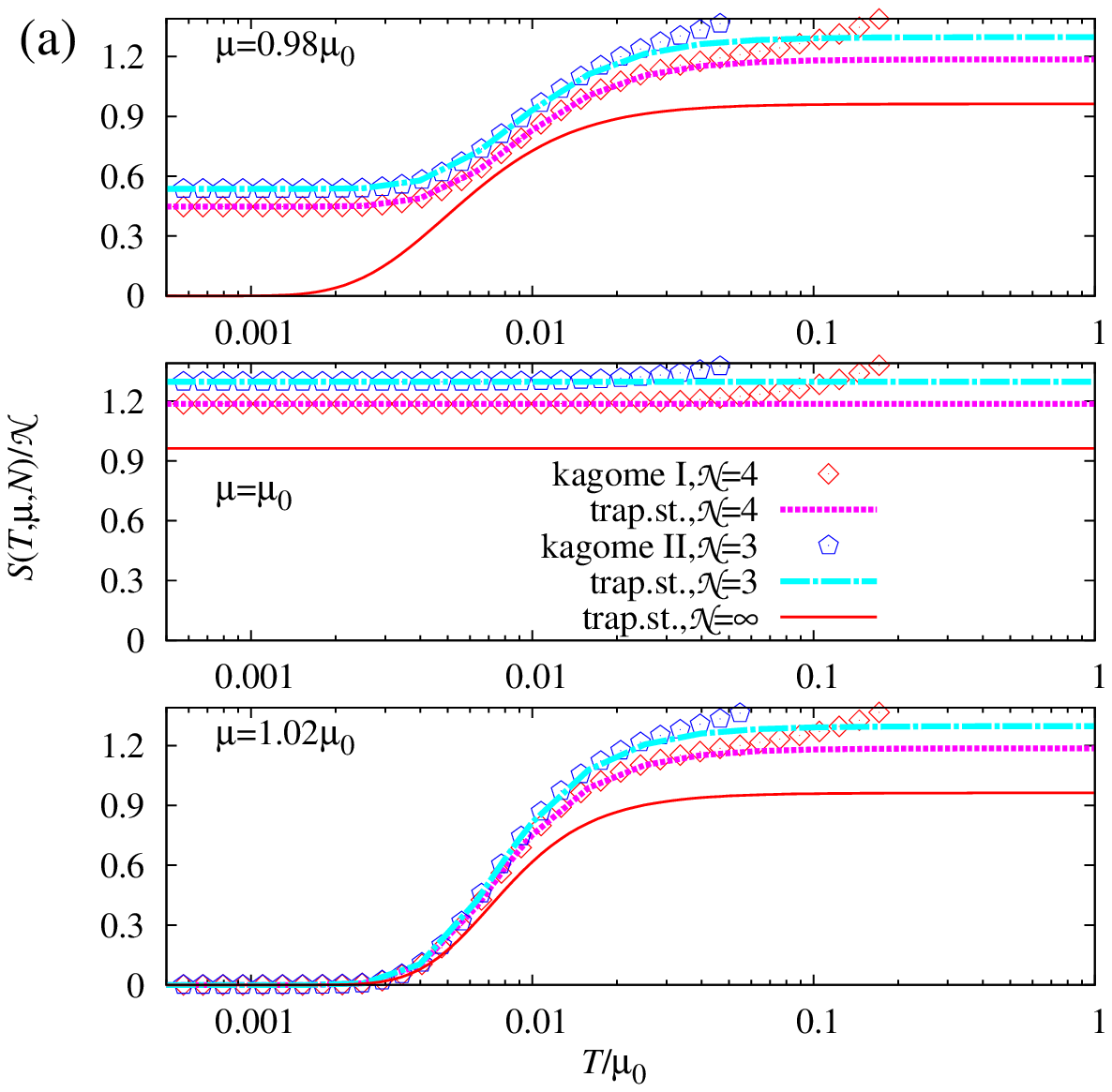}\\
\includegraphics[clip=on,width=\columnwidth,angle=0]{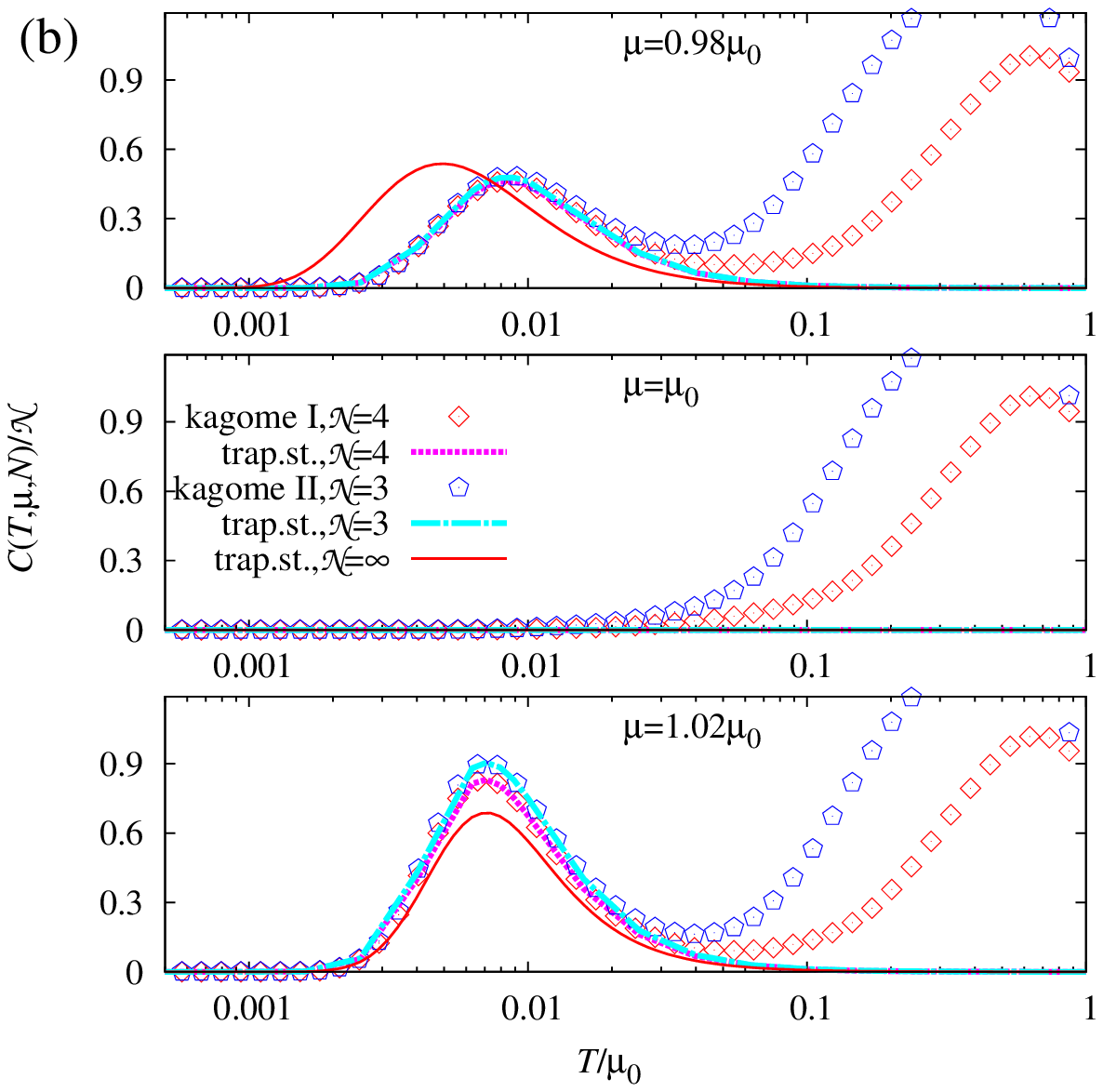}
\caption
{(Color online)
Entropy $S(T,\mu,N)/{\cal{N}}$
and
grand-canonical specific heat $C(T,\mu,N)/{\cal{N}}$
for the kagom\'{e} chain I ($N=12$, diamonds)
and
the kagom\'{e} chain II ($N=15$, the sectors with up to 7 electrons were taken into account, pentagons)
with $t=1$, $U\to\infty$.
(a)
$S(T,\mu,N)/{\cal{N}}$ versus temperature $T/\mu_0$ at $\mu=0.98\mu_0,\;\mu_0,\;1.02\mu_0$ (from top to bottom).
(b)
$C(T,\mu,N)/{\cal{N}}$ versus temperature $T/\mu_0$ at $\mu=0.98\mu_0,\;\mu_0,\;1.02\mu_0$ (from top to bottom).
We also show
the hard-dimer result for $N\to\infty$ as it follows from (\ref{5.06})
(thin solid lines)
as well as
the finite-size results
for ${\cal{N}}=3$ (dash-dotted lines) and ${\cal{N}}=4$ (dotted lines)
which follow from Eq.\ (\ref{5.05}) and include the contribution
of the leg states.}
\label{fig05}
\end{center}
\end{figure}

In Figs.\ \ref{fig04} and \ref{fig05} we show exact diagonalization data
for the temperature dependence of the entropy per cell $S(T,\mu,N)/{\cal{N}}$
at $\mu=0.98\mu_0,\;\mu_0,\;1.02\mu_0$
[panels (a)]
and
the temperature dependence of the grand-canonical specific heat per cell $C(T,\mu,N)/{\cal{N}}$
at $\mu=0.98\mu_0,\;\mu_0,\;1.02\mu_0$
[panels (b)]
for
the sawtooth chain ($N=12$) and the kagom\'{e} chain I ($N=9$) (Fig.\ \ref{fig04})
and
for the kagom\'{e} chain I ($N=12$) and the kagom\'{e} chain II ($N=15$) (Fig.\ \ref{fig05})
with $U\to\infty$.
We also report analytical predictions for finite ${\cal{N}}$ [obtained from
Eqs.\ (\ref{5.03}) and (\ref{5.05})] as well for infinite ${\cal{N}}$
[obtained from Eq.~(\ref{5.06})].

From Fig.\ \ref{fig03} we see
that in the  ground state
the average number of electrons per cell $\overline{n}(T,\mu,N)/{\cal{N}}$
as a function of $\mu/\mu_0$
exhibits a jump at $\mu/\mu_0=1$. 
This jump is related to the fact that all
ground states with $0\le n\le n_{\max}$ are degenerate at $\mu=\mu_0$. For $N
\to \infty$ the jump is  of height unity.
For $\mu/\mu_0<1$ a plateau appears, see Fig.\ \ref{fig03}.
The plateau width
(i.e., the charge gap)
increases as $U$ increases and rapidly approaches a saturation value
for all three finite lattices
(for the finite sawtooth chains with $N=12,\,16,\,20$
such data were reported in Fig.\ 1b of Ref.\ \onlinecite{dhr2007}).
What happens with the plateau width 
as $N$ increases?
For the sawtooth chain the plateau width 
is almost independent of $N$
(compare the results for $N=12,\,16,\,20$ in Fig.\ 1b of Ref.\ \onlinecite{dhr2007}).
For $U\to \infty$ we can also find the exact ground state for
$n=n_{\max}+1$, see Sec.~\ref{mag3}, which allows  to determine   
the size-independent plateau width $\Delta \mu =2t$.
By contrast, for the kagom\'{e} chains the plateau disappears as $N\to\infty$.

From Figs.\ \ref{fig04} and \ref{fig05} (temperature dependences of entropy and specific heat) we see
that analytical predictions for finite ${\cal{N}}$ as they follow from Eqs.\ (\ref{5.03}) and (\ref{5.05})
perfectly reproduce the exact diagonalization data at low temperatures.
Note, however,
that 
for finite $N$ the deviation from the hard-dimer description  is noticeable,
although in the thermodynamic limit $N\to\infty$ Eqs.\ (\ref{5.03}) and (\ref{5.05})
imply the one-dimensional hard-dimer behavior.
Finite-size effects are clearly seen in the panels corresponding to $\mu=0.98\mu_0$ (Figs.\ \ref{fig04} and \ref{fig05})
and to $\mu=\mu_0$ and $\mu=1.02\mu_0$ (Fig.\ \ref{fig05}).
The most prominent features seen in these plots are
a finite value of the entropy at very low temperatures for $\mu = \mu_0$
and
an extra low-temperature maximum in the specific heat $C(T,\mu,N)$ for $\mu \ne \mu_0$.
The value of the residual entropy for finite systems
as it follows from analytical predictions based on Eqs.\ (\ref{5.03}), (\ref{5.05})
and exact diagonalization are in perfect agreement.
A high-temperature maximum of $C(T,\mu,N)/{\cal{N}}$ around $T\approx \mu_0$
is common for any system with a finite band-width,
whereas a low-temperature peak of $C(T,\mu,N)/{\cal{N}}$ around $T\approx 0.01\mu_0$ for $\vert\mu-\mu_0\vert=0.02\mu_0$
emerges due to the manifold of localized electron states.
Indeed, the analytical predictions which follow from Eqs.\ (\ref{5.03}), (\ref{5.05}) for finite ${\cal{N}}$
are indistinguishable from the $U\to\infty$
exact diagonalization data in the low-temperature peak region.
We should, however, mention that for $N \to \infty$ the excitation
spectrum of the full Hubbard model
in the many-electron sectors is most likely gapless, in particular for
finite values of $U$ (compare Sec.~\ref{sec:IVC} and Ref.~\onlinecite{t-j}b). 
Such gapless excitations
could give rise to quantitative corrections at all temperatures
for $\mu < \mu_0$. Indeed such quantitative deviations are visible
for instance on finite-size sawtooth chains with $\mu=0.98\,\mu_0$
and $U=4\,t$.\cite{dhr2007,t-j}

\begin{figure}
\begin{center}
\includegraphics[clip=on,width=\columnwidth,angle=0]{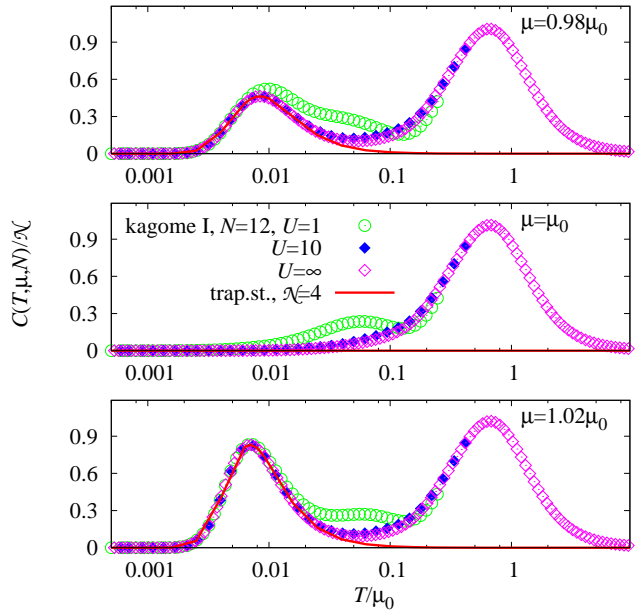}
\caption
{(Color online)
$C(T,\mu,N)/{\cal{N}}$ versus $T/\mu_0$
for the kagom\'{e} chain I with $N=12$ sites and $t=1$
for different values of $U$
[$U=1$ (empty circles),
$U=10$ (filled diamonds),
$U\to\infty$ (empty diamonds)].
The results for finite $U$ were obtained taking into account the sectors with up to 8 electrons.
We also show the result which follows from Eq.\ (\ref{5.05})
for ${\cal{N}}=4$ (lines).}
\label{fig06}
\end{center}
\end{figure}

To demonstrate the effect of finite $U$ on temperature dependences
we consider as an example the kagom\'{e} chain I with $N=12$ sites
and show $C(T,\mu,N)/{\cal{N}}$ versus $T/\mu_0$ for
$U=1,\;10,\;\infty$ (see Fig.\ \ref{fig06}).
For finite $U$ the calculation of thermodynamic quantities becomes
very time consuming and therefore we report the contribution of
the subspaces with up to a certain number of electrons which is
less than $2N$ thus restricting these data to not too high
temperatures. As can be seen from these data, the features at
sufficiently low temperature do not depend on the value of $U>0$.
Note, however, that at $U=1$ there are some visible corrections
down to temperatures $T/\mu_0 = {\cal O}(10^{-2})$.

\begin{figure}
\begin{center}
\includegraphics[clip=on,width=\columnwidth,angle=0]{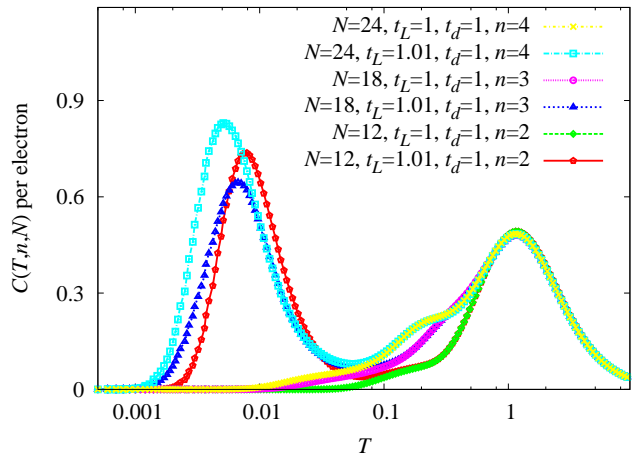}
\caption
{(Color online)
Canonical specific heat
$C(T,n,N)/n$ versus $T$
for
ideal ($t_d=t_L=1$)
and
distorted ($t_d=1$ and $t_L=1.01$)
kagom\'{e} chains I
with $n=N/6$ electrons;
$U\to\infty$.}
\label{fig07}
\end{center}
\end{figure}

From the experimental point of view it is important to discuss 
the stability of the features determined by localized electron states
with respect to small deviations from the ideal lattice geometry.
Then the conditions for the existence of flat bands are violated, 
and, as a result,
the exact degeneracy of the ground states in the subspaces 
with $n=1,\ldots,n_{\max}$ electrons is lifted.
However,
we are still faced with a set of a large number of low-lying energy levels which 
may dominate low-temperature thermodynamics
as $\mu$ is around $\mu_0$.
Deviations from ideal geometry for the sawtooth chain
(i.e., when $t^{\prime}$ slightly departs from the value $\sqrt{2}t$)
has been discussed already in Refs.~\onlinecite{honecker_richter}b and \onlinecite{t-j}b.
Below we consider the kagom\'{e} chain I
assuming the hoppings along legs to be slightly different from the hoppings along diamonds.
Specifically,
we put
$t_d=1$ along the diamonds
and
$t_L=1.01$ along the two legs.
In Fig.\ \ref{fig07} we report the results for the temperature dependence
of the canonical specific heat $C(T,n,N)/n$
for $n={\cal{N}}/2=N/6$ electrons.
For the ideal lattice $C(T,n,N)=0$ if $n\le n_{\max}$ in the low-temperature regime (Fig.\ \ref{fig07}).
A deviation from exact degeneracy
($t_d=1\ne t_L=1.01$)
produces a low-temperature peak (Fig.\ \ref{fig07}). 
This peak indicates a
separation of two energy scales, 
one is related to the manifold of low-lying trapped states, 
and the other one to the ordinary extended states. 
The position of this
extra peak depends on $t_d-t_L$, for instance for $t_d=1\ne t_L=1.1$ and
$N=24$ the
peak is at $T = 0.045$.
Although there are finite-size effects in the height and the position of the peak
its existence is not questioned.
We notice that the temperature dependence of the specific heat $C(T,n,N)/n$ is
an experimentally accessible quantity
and its well-pronounced low-temperature features
conditioned by localized electron states for small deviations from ideal lattice geometry
may increase chances to observe localized electron state effects.

\section{Magnetic properties}
\label{sec6}
\setcounter{equation}{0}

The study of ground-state magnetic properties of the Hubbard model (\ref{1.01}) on the considered lattices
is of great interest
and has been discussed since the early 1990s.
Thus, the sawtooth chain belongs to the one-dimensional version of Tasaki's lattice.
H.~Tasaki proved that the ground state of the sawtooth-chain Hubbard model is ferromagnetic and unique
[up to the trivial SU(2) degeneracy]
if the number of electrons $n=N/2$
(saturated ground-state ferromagnetism),
see Ref.\ \onlinecite{tasaki}a.
Later on, numerical studies of sawtooth chains of up to $N=12$ sites 
by Y.~Watanabe and S.~Miyashita\cite{watanabe}
revealed ground-state ferromagnetism (saturated and nonsaturated) for other values of $n$.
Moreover,
it was shown in Ref.\ \onlinecite{dhr2007} that within the localized electron picture
the model exhibits full polarization in the ground state for $n=N/2-1$ electrons
(only one-cluster states constitute the set of ground states)
and $60\%$ of the full polarization in the ground state for $n=N/2-2$ electrons if $N\to \infty$.

The two considered kagom\'{e} chains are line graphs.
This has been pointed out explicitly for the kagom\'{e} chain I,\cite{KMTT09}
where this connection has been known for a while,\cite{MSP03} albeit using a different
terminology.
According to a general theory elaborated by A.~Mielke
the repulsive Hubbard model on these lattices
should have ferromagnetic ground states for the number of electrons
$n \le N/3$ [periodic odd-${\cal{N}}$ kagom\'{e} chain I (nonbipartite parent graph)],
$n \le N/3+1$ [periodic even-${\cal{N}}$ kagom\'{e} chain I (bipartite parent graph)],
or
$n \le N/5+1$ [periodic kagom\'{e} chain II (bipartite parent graph)]
(Theorem of Ref.\ \onlinecite{mielke}a
or
Theorem 1 of Ref.\ \onlinecite{mielke}b).
Moreover,
according to A.~Mielke the ferromagnetic ground state is unique
apart from degeneracy due to SU(2) invariance
(i.e., saturated ground-state ferromagnetism),
if
$n=N/3$ (periodic odd-${\cal{N}}$ kagom\'{e} chain I),
$n=N/3+1$ (periodic even-${\cal{N}}$ kagom\'{e} chain I),
or
$n=N/5+1$  (periodic kagom\'{e} chain II)
(Theorem 2 of Ref.\ \onlinecite{mielke}b).

To reveal the ferromagnetic ground states we consider the operator
\begin{eqnarray}
\frac{\bm{S}^2}{{\cal{N}}^2}=\frac{\frac{1}{2}\left(
S^+\,S^- + S^-\,S^+\right)+{S^z}^2}{{\cal{N}}^2}
\label{6.01}
\end{eqnarray}
with the operators $S^\alpha$ defined in (\ref{eq:defS}).
The average at $T=0$, $\langle \bm{S}^2 \rangle_n/{\cal{N}}^2$
is given by the equal-weight average over all degenerate ground states 
for the given number of electrons $n$.
It satisfies
\begin{eqnarray}
0\le\frac{\langle \bm{S}^2\rangle_n}{{\cal{N}}^2}\le\frac{S_{\max}(S_{\max}+1)}{{\cal{N}}^2},
\;\;\;
S_{\max}=\frac{n}{2}.
\label{6.02}
\end{eqnarray}
If $\langle \bm{S}^2\rangle_n/{\cal{N}}^2$ achieves its maximal value
the ground state in the subspace with $n$ electrons is the saturated ferromagnetic ground state.
If $\langle \bm{S}^2\rangle_n/{\cal{N}}^2$ has a nonzero value which is less than the maximal value
the ground state in the subspace with $n$ electrons contains ferromagnetic ones
and is a nonsaturated ferromagnetic ground state.
To examine ground-state magnetism -- in particular the
existence of ferromagnetism -- for thermodynamically large systems
one has to consider the limit
${\cal{N}}\to\infty$, $n\to\infty$ preserving $n/{\cal{N}}=\rm{const}$.

Below we use the constructed many-electron ground states for $n=1,\ldots,n_{\max}$
to discuss systematically ground-state ferromagnetism at electron densities
$n\le n_{\max}$.
We complete our analytical arguments by numerics for finite systems.
Our results are consistent with general theorems of Tasaki and Mielke
for $n=n_{\max}$ and go beyond considering $n<n_{\max}$. 
Moreover,
we also report numerics for higher electron densities $n>n_{\max}$.

We start with a brief overview of our findings: (i) All 
ground states are fully polarized (saturated ferromagnetism) for $n={\cal N}$
and $n={\cal N}-1$ (all chains) and, in addition, 
for $n={\cal N}+1$
for the kagom\'{e} chain II 
and the even-${\cal N}$ kagom\'{e} chain I.
(ii) For smaller $n<{\cal N}-1$ the ground-state manifold contains fully and
partially polarized as well as paramagnetic states. Thus we have $\langle
\bm{S}^2\rangle_n < S_{\max}(S_{\max}+1)$. The magnetic polarization
$\langle
\bm{S}^2\rangle_n < S_{\max}(S_{\max}+1)$ decays monotonically with
increasing ${\cal N}-n$, $n<{\cal N}$, see Figs.~\ref{fig08} and
\ref{fig10}. For large ${\cal{N}}$ the decay becomes very rapid.    
(iii) While for finite systems there is a finite region of electron density 
$n/{\cal N}$ where ground-state ferromagnetism exists, this region shrinks
to one parameter point $n/{\cal N}=1$  for ${\cal N} \to \infty$. 
(iv) For $n/{\cal N} < 1$ the system shows Curie-like behavior with a
uniform zero-field magnetic susceptibility $\chi \propto T^{-1}$.

In what follows, we first discuss separately
the case of the systems with hard-dimer ground states only 
(see Sec.~\ref{mag1}) and
the case of the systems with ground states which also involve two-leg  states
(see Sec.~\ref{mag2}).
In Sec.~\ref{mag3} we report exact diagonalization data for finite systems
at higher electron densities $n>n_{\max}$.
We complete our discussion considering the low-temperature behavior
of the uniform zero-field magnetic susceptibility for $n\le n_{\max}$
in Sec.~\ref{mag4}.

\subsection{Hard-dimer ground states and ground-state magnetism}
\label{mag1}

To calculate $\langle \bm{S}^2\rangle_n$ we may further elaborate the $3\times 3$ transfer-matrix technique.
First,
we use SU(2) invariance of the Hubbard model (\ref{1.01})
to write $\langle \bm{S}^2\rangle_n=3\langle {S^z}^2\rangle_n$.
Second,
we notice 
that the average over all degenerate ground states for a given number of electrons $n$ is
$\langle {S^z}^2\rangle_n ={\cal{N}}\sum_{j=0}^{{\cal{N}}-1}\langle S_0^z S_j^z\rangle_n$,
where $S_j^z$ is the $z$-component of the spin operator of the trap $j$.
The operator $S_j^z$ 
acting on the hard-dimer ground states yields 0 (empty trap),
$1/2$ (occupied trap with spin-up electron),
$-1/2$ (occupied trap with spin-down electron).
Next with the help of the $3\times 3$ transfer matrix we find the grand-canonical $zz$ correlation function
$\langle S_0^z S_j^z\rangle_z$
(the subscript $z$ denotes the activity)
yielding the required canonical $zz$ correlation function
$\langle S_0^z S_j^z\rangle_n$ (see Appendix A).
As a result we obtain $\langle \bm{S}^2\rangle_n$ for $n=1,\ldots,{\cal{N}}-1$.
We recall that
$\langle \bm{S}^2\rangle_{{\cal{N}}}=({\cal{N}}/2)({\cal{N}}/2+1)$.
An alternative computation based on a hard-dimer mapping (see Appendix A)
yields the same results for $\langle \bm{S}^2\rangle_n$, $n=1,\ldots,{\cal{N}}$.

\begin{figure}
\begin{center}
\includegraphics[clip=on,width=\columnwidth,angle=0]{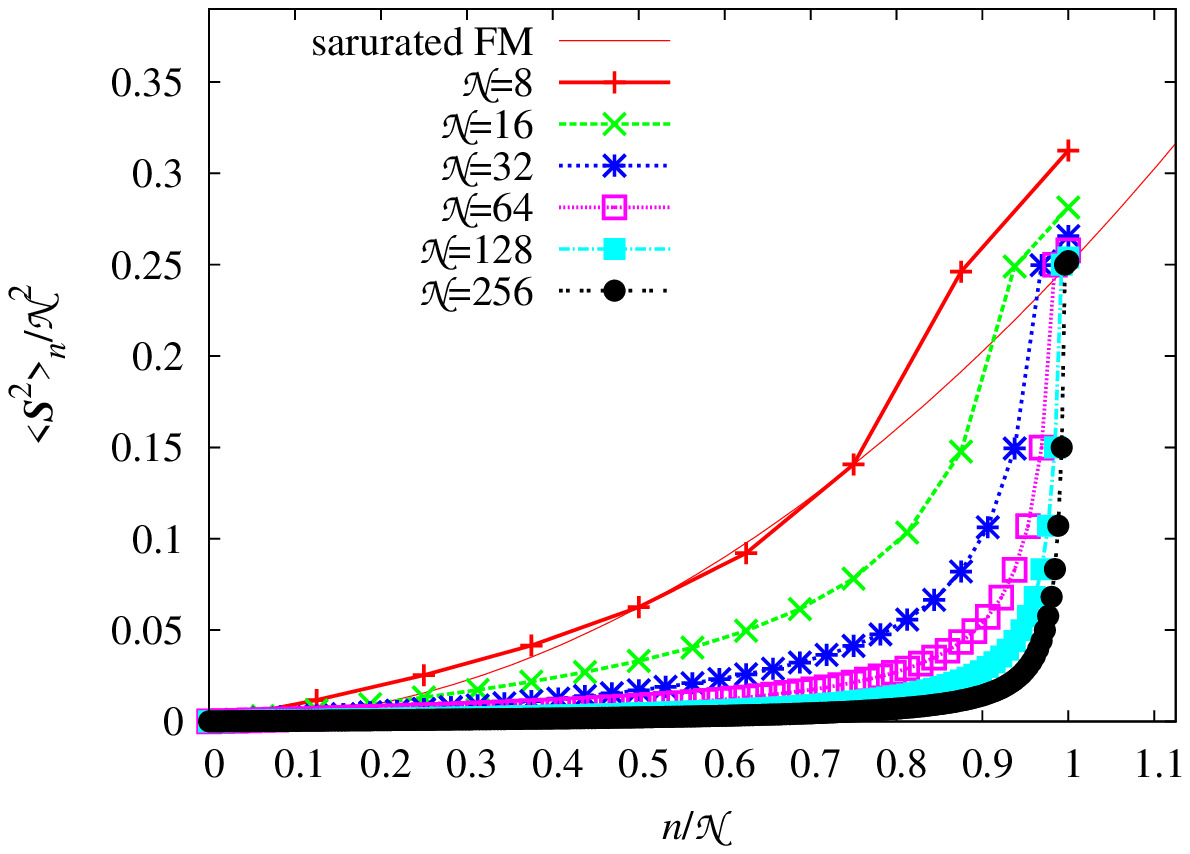}
\caption
{(Color online)
Average ground-state magnetic moment
of the sawtooth-Hubbard chain:
$\langle \bm{S}^2\rangle_n/{\cal{N}}^2$ versus $n/{\cal{N}}$
for ${\cal{N}}=8,\;16,\;32,\;64,\;128,\;256$.
For saturated ferromagnetic ground states
$\langle \bm{S}^2\rangle_n/{\cal{N}}^2$
would equal
$(n/{\cal{N}})^2/4$ (thin line)
in the thermodynamic limit ${\cal{N}}\to\infty$.}
\label{fig08}
\end{center}
\end{figure}

\begin{figure}
\begin{center}
\includegraphics[clip=on,width=\columnwidth,angle=0]{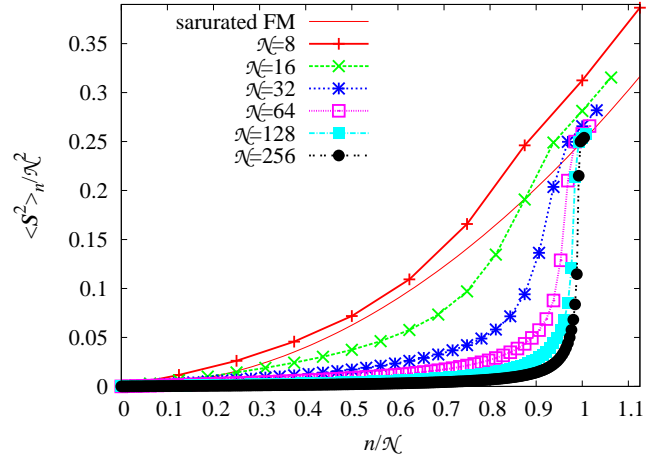}
\caption
{(Color online)
Average ground-state magnetic moment
of the kagom\'{e}-Hubbard chains:
$\langle \bm{S}^2\rangle_n/{\cal{N}}^2$ versus $n/{\cal{N}}$
for ${\cal{N}}=8,\;16,\;32,\;64,\;128,\;256$.
For saturated ferromagnetic ground states
$\langle \bm{S}^2\rangle_n/{\cal{N}}^2$
would equal
$(n/{\cal{N}})^2/4$ (thin line)
in the thermodynamic limit ${\cal{N}}\to\infty$.}
\label{fig10}
\end{center}
\end{figure}

Using the transfer-matrix method  we have 
calculated $\langle \bm{S}^2\rangle_n/{\cal{N}}^2$,
$n=1,\ldots,{\cal{N}}$ for systems
of up to  ${\cal{N}}=256$, 
see Fig.~\ref{fig08}. 
We obviously have $\langle \bm{S}^2\rangle_{n}=S_{\max}(S_{\max}+1)$
for $n={\cal{N}}$ and ${\cal{N}}-1$
independently of the system size as predicted above.
From the data shown in Fig.\ \ref{fig08}
we see clearly that $\langle \bm{S}^2\rangle_n/{\cal{N}}^2$ becomes smaller for any
fixed $n/{\cal{N}}$, $0< n/{\cal{N}}< 1$, as ${\cal{N}}$ increases
(compare the curves for ${\cal{N}}=8$ and ${\cal{N}}=256$). More precisely,
the curves shown in Fig.~\ref{fig08} suggest $\langle
\bm{S}^2\rangle_n/{\cal{N}}^2\to 0$ for ${\cal{N}} \to \infty$
at $n/{\cal{N}}=\mbox {const}$.
Note that for saturated ferromagnetic ground states
$\langle \bm{S}^2\rangle_n/{\cal{N}}^2$ would equal $(n/{\cal{N}})^2/4$
in the thermodynamic limit ${\cal{N}}\to\infty$. In Fig.~\ref{fig08},
one observes increasing differences between the curve $(n/{\cal{N}})^2/4$
and the data for $\langle \bm{S}^2\rangle_n/{\cal{N}}^2$ which
approaches zero for $n/{\cal{N}}<1$. 
We have performed
a finite-size extrapolation
of $\langle \bm{S}^2\rangle_n/{\cal{N}}^2$  for several fixed electron densities
$n/{\cal{N}}= 1/2,\;3/4,\;7/8$ to estimate  the corresponding values for
${\cal{N}}\to\infty$, and we find, indeed, $\lim_{{\cal N} \to \infty}\langle
\bm{S}^2\rangle_n/{\cal{N}}^2=0$ for those values of $n/{\cal{N}}$.

In a next step we will give analytical predictions for $\langle \bm{S}^2\rangle_n/{\cal{N}}^2$.
We first   
collect the values for  $\langle \bm{S}^2\rangle_n$  for system sizes up to
${\cal{N}}=8$ in Table~\ref{table4}. Note that the exact-diagonalization data
for $\langle \bm{S}^2\rangle_n$
for the sawtooth chain confirm the values given in Table~\ref{table4}.
\begin{table}
\begin{center}
\caption
{Ground-state $\langle \bm{S}^2\rangle_n$ for $n=1,\ldots,{\cal{N}}$ electrons
for the sawtooth-Hubbard chain.
\label{table4}}
\begin{tabular}{|c||c|c|c|c|c|c|c|c|}
\hline
             &$n=1$        &$n=2$          &$n=3$           &$n=4$         &$n=5$           &$n=6$&$n=7$        &$n=8$\\
\hline
\hline
${\cal{N}}=2$&$\frac{3}{4}$&$2$            &-               &-             &-               &-   &-             &-    \\
\hline
${\cal{N}}=3$&$\frac{3}{4}$&$2$            &$\frac{15}{4}$  &-             &-               &-   &-             &-    \\
\hline
${\cal{N}}=4$&$\frac{3}{4}$&$\frac{9}{5}$  &$\frac{15}{4}$  &$6$           &-               &-   &-             &-    \\
\hline
${\cal{N}}=5$&$\frac{3}{4}$&$\frac{12}{7}$ &$\frac{63}{20}$ &$6$           &$\frac{35}{4}$  &-   &-             &-    \\
\hline
${\cal{N}}=6$&$\frac{3}{4}$&$\frac{5}{3}$  &$\frac{81}{28}$ &$\frac{24}{5}$&$\frac{35}{4}$  &$12$&-             &-    \\
\hline
${\cal{N}}=7$&$\frac{3}{4}$&$\frac{18}{11}$&$\frac{11}{4}$  &$\frac{30}{7}$&$\frac{27}{4}$  &$12$&$\frac{63}{4}$&-    \\
\hline
${\cal{N}}=8$&$\frac{3}{4}$&$\frac{21}{13}$&$\frac{117}{44}$&$4$           &$\frac{165}{28}$&$9$ &$\frac{63}{4}$&$20$ \\
\hline
\end{tabular}
\end{center}
\end{table}
By inspecting the values of $\langle \bm{S}^2\rangle_n$ reported in Table \ref{table4}
we may notice that
\begin{eqnarray}
\langle \bm{S}^2\rangle_2 &=&
\frac{6+3({\cal{N}}-3)}{3+2({\cal{N}}-3)} 
\stackrel{{\cal{N}}\to\infty}{\longrightarrow}
\frac{3}{2},
\nonumber \\
\langle \bm{S}^2\rangle_3&=&
\frac{45+18({\cal{N}}-4)}{12+8({\cal{N}}-4)}
\stackrel{{\cal{N}}\to\infty}{\longrightarrow}
\frac{9}{4}, 
\nonumber \\
\langle \bm{S}^2\rangle_4&=&
\frac{54+18({\cal{N}}-5)}{9+6({\cal{N}}-5)}
\stackrel{{\cal{N}}\to\infty}{\longrightarrow}
3,
\nonumber \\
\langle \bm{S}^2\rangle_5&=&
\frac{210+60({\cal{N}}-6)}{24+16({\cal{N}}-6)}
\stackrel{{\cal{N}}\to\infty}{\longrightarrow}
\frac{15}{4}.
\label{6.03}
\end{eqnarray}
This leads to a guess for the thermodynamic limit: 
\begin{eqnarray}
\lim_{{\cal{N}}\to\infty}\langle \bm{S}^2\rangle_n
=\frac{3}{4}n,
\;\;\; 
\frac{n}{\cal{N}} < 1.
\label{6.06}
\end{eqnarray}
This result is in accordance with the numerical results presented in
Fig.~\ref{fig08},
and  gives again evidence for a paramagnetic ground state  in the infinitely large
system.

Summarizing the above analysis we conclude that there is no finite range
of ground-state ferromagnetism
for electron densities $n/{\cal{N}}<1$ as ${\cal{N}}\to\infty$.

\subsection{Ground states involving two-leg states and ground-state magnetism}
\label{mag2}

We turn now to the periodic kagom\'{e} chain I with an even number
of cells ${\cal{N}}$ and the kagom\'{e} chain II
for the number of electrons $n\le{\cal{N}}+1$.
As was explained in Sec.~\ref{sec4},
the states involving two-leg states 
increase the ground-state degeneracy for $n=1,\ldots,{\cal{N}}$ 
by the number $L_{\cal{N}}(n)$
[see Eq.\ (\ref{4.05})].
All these additional states
[except the one state for $n=2$, see Eq.\ (\ref{4.04})]
are fully polarized, i.e., $\bm{S}^2=(n/2)(n/2 +1)$.
Therefore, for the average value we have
$\langle \bm{S}^2\rangle_n \ge \langle \bm{S}^2\rangle_n\vert_{D}$, where $\langle \bm{S}^2\rangle_n\vert_{D}$ is
the corresponding value considering hard-dimer states only (i.e.,
it is that value considered in Sec. \ref{mag1}, 
see, e.g., Table \ref{table4}).
Recalling that $D_{{\cal{N}}}(n)$
is the number of hard-dimer ground states
one finds
\begin{eqnarray}
\langle \bm{S}^2\rangle_n
=
\frac{D_{{\cal{N}}}(n)}{D_{{\cal{N}}}(n)+L_{{\cal{N}}}(n)}\langle \bm{S}^2\rangle_n\vert_{D}
\nonumber\\
+
\frac{L_{{\cal{N}}}(n)-\delta_{n,2}}{D_{{\cal{N}}}(n)+L_{{\cal{N}}}(n)}\;\frac{n}{2}\left(\frac{n}{2}+1\right).
\label{6.07}
\end{eqnarray}
This formula is valid for $n=1,\ldots,{\cal{N}}-1$.
We recall that
$\langle \bm{S}^2\rangle_{{\cal{N}}}=({\cal{N}}/2)({\cal{N}}/2+1)$
and
$\langle \bm{S}^2\rangle_{{\cal{N}}+1}=[({\cal{N}}+1)/2][({\cal{N}}+1)/2+1]$.
Based on Eq.\ (\ref{6.07})
we again can calculate $\langle \bm{S}^2\rangle_n$ for the kagom\'{e} chains
of up to ${\cal N}=256$ cells, see Fig.~\ref{fig10}. Noticeable deviations
from  $\langle \bm{S}^2\rangle_n\vert_{D}$ appear only for small
${\cal{N}}$. 
Hence the data shown in Fig.~\ref{fig10} give clear evidence that the
contribution of the two-leg states becomes irrelevant for large systems.
This conclusion is supported by further inspection of Eq.~(\ref{6.07}).
Using Eq.~(\ref{6.06}) to replace $\langle \bm{S}^2\rangle_n\vert_{D}$ in the limit ${\cal{N}}\to\infty$ 
we can write Eq.~(\ref{6.07})
for large ${\cal N}$ as 
\begin{eqnarray}
\lim_{{\cal{N}}\to\infty}\langle \bm{S}^2\rangle_n
=R_n\;\frac{3}{4}n + (1-R_n)\;\frac{n}{2}\left(\frac{n}{2}-1\right),
\nonumber\\
R_n=\lim_{{\cal{N}}\to\infty}\frac{D_{{\cal{N}}}(n)}{D_{{\cal{N}}}(n)+L_{{\cal{N}}}(n)}.
\label{6.08}
\end{eqnarray}
According to  Eqs.\ (\ref{a.05}), (\ref{a.06}), (\ref{5.03}), and
(\ref{5.05}) 
the quantity $D_{{\cal{N}}}(n)$
is proportional to the $n$th derivative
of $\xi_{+}^{{\cal{N}}}+\xi_{-}^{{\cal{N}}}+\xi_3^{{\cal{N}}}$  with respect to $z$ at $z=0$,
whereas $L_{{\cal{N}}}(n)$
is proportional to the $n$th derivative
of $\xi_{4}^{{\cal{N}}}+\xi_{5}^{{\cal{N}}}+\xi_{6}^{{\cal{N}}}$  with respect to $z$ at $z=0$.
Evaluating derivatives at $z=0$ 
we find that for ${\cal{N}}\to\infty$ the ratio $L_{{\cal{N}}}(n)/D_{{\cal{N}}}(n)\propto 1/{\cal{N}}$
and $\lim_{{\cal{N}}\to\infty}L_{{\cal{N}}}(n)/D_{{\cal{N}}}(n)= 0$.
Evidently,
$R_n=1$ in Eq.\ (\ref{6.08}) implies that Eq.~(\ref{6.06}) is valid again,
i.e., the
ground state is paramagnetic for the infinitely large system.

\subsection{Results for $n> n_{\max}$}
\label{mag3}

\begin{table*}[t]
\begin{center}
\caption{Electron numbers $n_{\max} < n < N$ for which saturated ground-state ferromagnetism
in the limit  $U\to\infty$ exists. Note that for  $N>20$
the size of the Hamiltonian matrix becomes very large. Therefore, for $N>20$
only a few sectors $n>n_{\max}$ are accessible by numerical
calculation. 
In particular, 
for $N=24$ we can examine only the sectors with up to $n=16$ electrons
and for $N=30$ with up to $n=10$ electrons.
Note further that for 
$N=40$ and
$N=50$ we cannot  reach the sectors with $n> n_{\max}$.
\label{table_FM}}
\begin{tabular}{|c||c|c|c|c|}
\hline
chain         & ${\cal N}=4$                & ${\cal N}=6$              & ${\cal N}=8$           & ${\cal N}=10$ \\ 
\hline
\hline
sawtooth      & $N=8$: $n=5,7$              & $N=12$: $n=7,9,11$        & $N=16$: $n=9,11,13,15$ & $N=20$: $n=11,13,15,17,19$ \\
\hline
kagom\'{e} I  & $N=12$: $n=8,11$            & $N=18$: $n=9,10,14,15,17$ & $N=24$: $n=11,$ --     & $N=30$: --    \\
\hline
kagom\'{e} II & $N=20$: $n=7,8,10,13,17,19$ & $N=30$: $n=9,10,$ --      & $N=40$: --             & $N=50$: --    \\ 
\hline
\end{tabular}
\end{center}
\end{table*}

The flat-band ferromagnets discussed above may exhibit
ferromagnetic ground-state ordering even for electron numbers $n > n_{\max}$.
However, this
ground-state ferromagnetism
occurs only for sufficiently large $U$
and therefore is definitely different from the true flat-band ferromagnetism which emerges for any arbitrary small
$U$.

The occurrence of a saturated ferromagnetic ground state can be well
understood for the sawtooth chain with $n=n_{\max}+1$, $n_{\max}={\cal{N}}$.
For $U=0$ the sawtooth chain has two single-electron bands separated by the energy gap
$\Delta_1=\varepsilon_2(\pi)-\varepsilon_1=2t$, see Eq.\ (\ref{3.01}).
If $U$ is small (in comparison with, e.g., $\Delta_1$)
the ground state in the subspace with $n={\cal{N}}+1$ electrons
is a complicated many-body state.
However,
if $U$ is sufficiently  large $U>U_c({\cal{N}}+1)$
(and in particular in the limit $U\to\infty$)
it might be energetically favorable to avoid Hubbard repulsion. This can be 
realized  by occupying all trapping cells ({\sf V}-valleys) with ${\cal{N}}$,
say, spin-up electrons and putting the one remaining electron also with
$\sigma=\uparrow$ into the next (dispersive) band.
Indeed, it is easy to show that 
for the periodic even-${\cal{N}}$ sawtooth chain such a state 
\begin{eqnarray}
\vert\varphi_{{\cal{N}}+1}\rangle
=\alpha_{2,\pi,\uparrow}^{\dagger}
l_{0,\uparrow}^{\dagger}l_{2,\uparrow}^{\dagger}\ldots l_{N-2,\uparrow}^{\dagger}\vert 0\rangle
\label{6.09}
\end{eqnarray}
is a true eigenstate with the energy 
${\cal{N}}\varepsilon_1+\varepsilon_2(\pi)$.
The trapping-cell operators $l_{2j,\uparrow}^{\dagger}$ are defined
in Eq.~(\ref{3.02}) and $\alpha_{2,\pi,\uparrow}^{\dagger}$ creates an
electron in the dispersive band with $\kappa=\pi$ and $\sigma=\uparrow$.   
Other states belonging to a spin-$[({\cal{N}}+1)/2]$ SU(2) multiplet
can be obtained by applying $S^-$ to the state (\ref{6.09}).
This kind of saturated ground-state ferromagnetism for the sawtooth chain was
first found numerically 
by Y.~Watanabe and S.~Miyashita\cite{watanabe}
(see also Refs.\ \onlinecite{tanaka_idogaki,dhr2009}).

We may expect such fully polarized ferromagnetic ground states for sufficiently large $U$ for
further electron
numbers $n_{\max}<n<N$,
where $n=N-1$ is that electron number where the well-known 
Nagaoka theorem\cite{nagaoka} holds.
Using Lanczos exact diagonalization of finite systems we have 
investigated this question
for the sawtooth and the kagom\'{e} chains I and II.
We list our numerical findings  for $U\to\infty$ in Table~\ref{table_FM}. Indeed, fully
polarized ferromagnetic ground states exist for various electron numbers $n$ in
the range $n_{\max} < n < N$.
For other values not listed in Table~\ref{table_FM} the ground state is either partially polarized,
i.e., $0 < \langle \bm{S}^2\rangle_n < S_{\max}(S_{\max}+1)$, 
or it is a singlet, i.e., $\langle \bm{S}^2\rangle_n=0$, with spiral structure, see
also Ref.~\onlinecite{watanabe}.
However, a detailed discussion of this issue 
goes beyond the scope of the present paper.

\subsection{Low-temperature behavior of the magnetic susceptibility} 
\label{mag4}

We complete our discussion of the magnetic properties with a brief
consideration of the low-temperature behavior of the uniform zero-field magnetic susceptibility
$\chi$.   
Using the standard arguments for deriving the uniform zero-field Langevin susceptibility
we may write the trapped-state contribution to $\chi$ as  
\begin{eqnarray}
\chi(T,n,N) 
=\frac{\langle \bm{S}^2\rangle_n}{3T}
\label{chi_n}
\end{eqnarray}
with $\langle \bm{S}^2\rangle_n$ calculated in Secs.\ \ref{mag1} and \ref{mag2}.
Thus we may expect a Curie-like behavior of the susceptibility of the considered Hubbard chains 
at low temperatures in case of paramagnetic ground states.
In the thermodynamic limit we have  $\lim_{{\cal{N}}\to\infty}\langle
\bm{S}^2\rangle_n =3n/4$ in the paramagnetic region ($n/{\cal N}<1$), see Eq.~(\ref{6.06}). 
Therefore the Curie constant is $n/4$, which corresponds to a system of $n$ independent spins 1/2.
Note, however, that in case of a ferromagnetic ground state the low-temperature dependence
of $\chi$ is expected to be different.
Thus for the spin-1/2 ferromagnetic Heisenberg chain 
the Bethe ansatz 
yields $\chi\propto T^{-2}$ (see Ref.~\onlinecite{yamada}) 
which holds also for weakly frustrated chains.\cite{haertel}

\begin{figure}
\begin{center}
\includegraphics[clip=on,width=\columnwidth,angle=0]{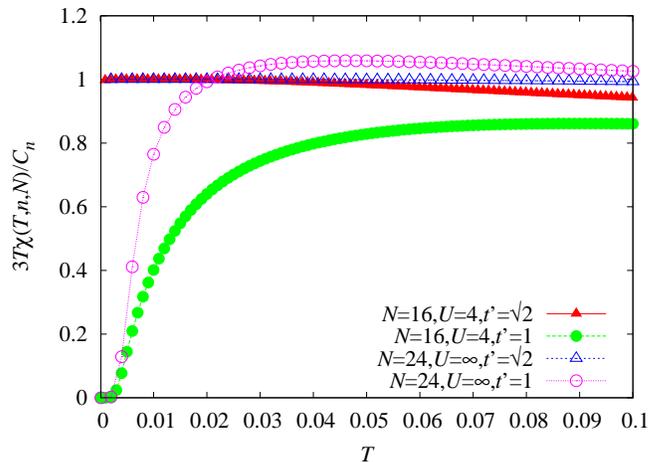}
\caption
{(Color online)
Uniform magnetic susceptibility $3T\chi(T,n,N)/C_n$ for the sawtooth chain ($t=1$)
with $n=4$ electrons 
and $N=16$ sites with $U=4$ (filled triangles and circles) 
as well as $N=24$ sites with $U \to\infty$ (empty triangles and circles).
Triangles correspond to $t^{\prime}=\sqrt{2}$
and
circles correspond to $t^{\prime}=1$.
Here we use for the normalization of the vertical axis
$C_n=4$ for ${\cal{N}}=8$ and $C_n=60/17$ for ${\cal{N}}=12$.}
\label{fig12}
\end{center}
\end{figure}

Again we have confirmed Eq.~(\ref{chi_n}) by numerical calculations for finite systems.
As an example some results for the sawtooth chain  of two different
lengths are presented in Fig.~\ref{fig12} for $t^{\prime}=\sqrt{2}$
and $t^{\prime}=1$.
From Fig.~\ref{fig12}
it is obvious that for the sawtooth chain with a flat band, i.e.,
$t^{\prime}=\sqrt{2}$, Eq.~(\ref{chi_n}) holds at low temperatures.
Moreover, the obtained Curie constants imply 
$\langle \bm{S}^2\rangle_4=4$ for ${\cal{N}}=8$
and
$\langle \bm{S}^2\rangle_4=60/17$ for ${\cal{N}}=12$
in agreement with calculations of Sec.~\ref{mag1}.
The temperature region where this relation is
valid increases with growing $U$. 
In case of a dispersive lowest band, i.e., $t'=1$, the quantity $\langle \bm{S}^2\rangle_n$ depends on $U$ and $n$. 
Indeed, it has been observed previously\cite{penc}
that in the sawtooth chain with $t'=t$ for quarter filling and less than quarter filling
the Coulomb repulsion $U$ may drive transitions
from  singlet ground states present at small values of $U$
to ferromagnetic ground states present at large values of $U$.
For the values of $U$ and $n$ considered in Fig.~\ref{fig12} we have
singlet ground states, 
i.e., $\langle \bm{S}^2\rangle_n=0$,
for $t^{\prime}=1$. Hence, contrary to the flat-band case   
we have $T\chi(T,n,N)=0$ as $T\to 0$.

\section{Relation to the $XXZ$ model} 
\label{xxz}

Finally we want discuss the relation between the
exact many-electron states considered in this paper and the localized magnon
states found for the $XXZ$ Heisenberg
antiferromagnet.\cite{localized_magnons,loc_mag_thermo1,loc_mag_thermo2}

First we notice 
that the localized magnon states for the $XXZ$ Heisenberg antiferromagnet
on all three lattices
can be also mapped onto the model of hard dimers on an auxiliary simple chain.
However,
in contradistinction to the electron model,
the localized magnons cannot sit in neighboring traps
and there is only one possibility to occupy a trap. Moreover, 
for the
kagom\'{e} chains the simultaneous occupation of a leg and a diamond or
hexagon trapping cell 
by magnons is not allowed.
Hence, we are faced with an example where the Pauli principle leads to less
constraints in comparison to hard-core bosonic systems.  
As a result, the number of states for the $XXZ$ model is given by
hard dimers on a chain of only ${\cal{N}}$ sites instead of the $2{\cal{N}}$
sites as is the case for the Hubbard
model.\cite{loc_mag_thermo1,loc_mag_thermo2}

Further differences can be emphasized between the localized magnons and the localized electrons
for these three chains.
First, in the subspace with $n={\cal{N}}/2$ magnons the ground-state degeneracy
equals 2
whereas in the corresponding subspace with $n={\cal{N}}$ electrons the hard-dimer ground-state degeneracy
equals ${\cal{N}}+1$.
For the antiferromagnetic Heisenberg kagom\'{e} chains
the extended single-magnon states (two-leg states) also appear,
however, because of the stronger constraint 
only in the subspaces with $n=1$ and $n=2$ magnons.
This increases the total ground-state degeneracy at the
saturation magnetic field $h=h_{{\rm{sat}}}$ exactly by 2
(compare Ref.~\onlinecite{loc_mag_thermo2}b).
By contrast,
for the Hubbard kagom\'{e} chains
the extended single-electron states (two-leg states) appear in all subspaces with
$n=1,\ldots,{\cal{N}}+1$ electrons
thus noticeably increasing the ground-state degeneracy at $\mu=\mu_0$.

Despite the differences between the fermionic and the hard-core bosonic systems
stressed above, the existence of localized states in both systems leads to
some common features. For instance,  the jump in the number of electrons as
a function of the chemical potential $\mu$ at $\mu=\mu_0$ found
for the electron system  
corresponds to  
a jump in the magnetization curve as a function of the external magnetic field at 
the saturation field $h=h_{{\rm{sat}}}$ seen in the $XXZ$ model.
Moreover, in both cases the contribution of the localized states to the partition function
can be calculated explicitly by a transfer-matrix method, 
which leads to simple analytical expressions
for the low-temperature behavior of various thermodynamic quantities, such 
as the entropy or the specific heat.

\section{Conclusions}
\label{sec7}
\setcounter{equation}{0}

In this paper,
we have considered two different types of flat-band ferromagnets,
namely,
the sawtooth Hubbard chain (Tasaki's model)
and
two kagom\'{e} Hubbard chains (Mielke's models).
For these three models we have constructed the complete set of ground states for electron
numbers
$n\le n_{\max}$, where
$n_{\max}={\cal{N}}$
(sawtooth, odd-${\cal{N}}$ kagom\'{e} I)
or
$n_{\max}={\cal{N}}+1$
(even-${\cal{N}}$ kagom\'{e} I, kagom\'{e} II)
with ${\cal{N}}=N/2$ (sawtooth), ${\cal{N}}=N/3$ (kagom\'{e} I), and ${\cal{N}}=N/5$ (kagom\'{e} II).
In these ground states the electrons are trapped on restricted areas of the
full lattice.  
Using a transfer-matrix method, see  Appendix~\ref{a},  
we have calculated the degeneracy of the ground states $g_{{\cal{N}}}(n)$. 
The ground-state degeneracy grows
rapidly with increasing system size, and a finite residual entropy per site remains
in the thermodynamic limit. 
Moreover,
we have calculated exactly the contribution of the highly degenerate ground-state
manifold to the partition function.  The low-temperature thermodynamics 
around a particular value of the chemical potential $\mu_0$
($\mu_0=2t$ for all three models) is dominated by these trapped ground
states leading to a low-energy scale separated from the usual energy scale
determined by the band width.  
In the thermodynamic limit $N\to\infty$ all models exhibit identical
thermodynamic behavior in this regime
which is analogous to that of classical one-dimensional hard dimers.

The trapped ground states lead also to particular magnetic behavior
including ground-state ferromagnetism as well as paramagnetic behavior.    

Moreover, with this study we have (i) 
illuminated relations between 
Tasaki's and Mielke's flat-band ferromagnets, and (ii)
the relations
between frustrated quantum Heisenberg antiferromagnets 
and the Hubbard flat-band ferromagnets.
In spite of some similarities in the mathematical description of both correlated quantum lattice systems
owing to localized one-particle states,
the elaboration of a comprehensive theory for the Hubbard flat-band ferromagnet
in higher dimensions using some ideas from localized-spin systems  
remains an unsolved problem and calls for further efforts.

\section*{Acknowledgments}

The numerical calculations were performed using J.~Schulenburg's {\it spinpack}.
Financial support of the DFG is gratefully acknowledged
(projects Ri615/16-1 and Ri615/18-1
and a Heisenberg fellowship for A.H.\ under project HO~2325/4-1).
O.D.\ acknowledges the kind hospitality
of the MPIPKS-Dresden in 2006 and 2009
and
of the University of Magdeburg in the autumn of 2008 and 2009.

\appendix
\section{Transfer-matrix counting of localized hard-dimer electron states}
\label{a}

The ground states of electrons localized on trapping cells 
({\sf V}-valleys, diamonds or hexagons, respectively) for $U>0$
can be counted using a transfer-matrix method.\cite{baxter}
In Ref.~\onlinecite{dhr2007} we have associated two sites to each trapping
cell and used a hard-dimer mapping for this counting. Here we will
present a direct solution of the same problem using a $3 \times 3$
transfer matrix.

Recall that the ground states of the considered systems,
i.e., the sawtooth chain or the odd-${\cal{N}}$ kagom\'{e} chain I with $\mu$ around $\mu_0$,
can be obtained by populating the trapping cells with electrons with spin-up and
spin-down according to the following rules:
1) Each trap may be empty, occupied by one spin-up electron or occupied by one spin-down electron.
That is, for each trap $j=0,1,\ldots,{\cal{N}}-1$
we have three trap states, $s_j=0,\uparrow,\downarrow$.
A ground state of the chain can be thought of as a certain sequence of the trap states.
2) Moving along the successive traps
(the choice of the first trap is totally arbitrary,
e.g., we may take $j=0$)
we must allow only {\em {one}} sequence of two trap states
corresponding to the neighboring cells being occupied by differently polarized electrons.
This is only a convention for the correct counting of the number of the ground states.
(Note, however, that the convention does not work for the number of electrons $n={\cal{N}}$
yielding only 2 states instead of correct number ${\cal{N}}+1$.)
For instance,
let us allow the sequence of trap states $s_j=\uparrow$, $s_{j+1}=\downarrow$
and forbid the sequence of trap states $s_j=\downarrow$, $s_{j+1}=\uparrow$.

These rules can be encoded with a transfer matrix\cite{another}
\begin{eqnarray}
{\bf{T}}&=&
\left(
\begin{array}{ccc}
T(0,0)          & T(0,\uparrow)          & T(0,\downarrow)         \\
T(\uparrow, 0)  & T(\uparrow,\uparrow)   & T(\uparrow,\downarrow)  \\
T(\downarrow,0) & T(\downarrow,\uparrow) & T(\downarrow,\downarrow)
\end{array}
\right)
\nonumber \\
&=&
\left(
\begin{array}{ccc}
1 & 1 & 1 \\
z & z & z \\
z & 0 & z
\end{array}
\right)
\label{c.01}
\end{eqnarray}
with the activity $z=\exp(-\varepsilon_1/T)$.
Now we can write down the contribution
of all allowed sequences $s_0,\ldots,s_{{\cal N}-1}$
to the grand-canonical partition function $\Xi_{{\trap}}(z,{\cal{N}})$ (\ref{5.02}) as follows
\begin{eqnarray}
\Xi_{{\trap}}(z,{\cal{N}})
&=&{\rm{Tr}}{\bf{T}}^{{\cal{N}}}
\nonumber\\
&=&\xi_+^{{\cal{N}}}+\xi_-^{{\cal{N}}}+\xi_0^{{\cal{N}}},
\nonumber\\
\xi_{\pm}&=&\frac{1}{2}+z\pm\sqrt{\frac{1}{4}+z},
\quad
\xi_0=0.\quad
\label{c.02}
\end{eqnarray}
The hard-dimer computation yields the alternative representation
$\Xi_{{\trap}}(z,{\cal{N}}) = \lambda_+^{2{\cal{N}}}+\lambda_-^{2{\cal{N}}}$
where $\lambda_{\pm}=1/2\pm\sqrt{1/4+\exp x}$ are the eigenvalues
of a $2\times2$ transfer matrix.\cite{dhr2007}
Since $\xi_{\pm}=\lambda_{\pm}^2$
and $\xi_0=0$, the two expressions are in fact equivalent.

We turn now to the canonical description. 
More specifically, we consider $n\le {\cal{N}}$ trapped states
on a (periodic) chain with ${\cal{N}}$ cells.
We are interested in the canonical partition function of such a system
${\cal{Z}}(n,{\cal{N}})$ which counts the number of spatial configurations
of $n$ trapped electrons.
Using the relation between the canonical partition function and the grand-canonical partition function
\begin{eqnarray}
\label{a.05}
\Xi_{{\trap}}(z,{\cal{N}})
=
\sum_{n=0}^{{\cal{N}}}z^n{\cal{Z}}(n,{\cal{N}})
\end{eqnarray}
we immediately find that
\begin{eqnarray}
\label{a.06}
{\cal{Z}}(n,{\cal{N}})
=\frac{1}{n!}\left.\frac{d^n\Xi_{{\trap}}(z,{\cal{N}})}{dz^n}\right\vert_{z=0}.
\end{eqnarray}
After simple (but becoming tedious as $n$ increases) calculations we get
\begin{widetext}
\begin{eqnarray}
{\cal{Z}}(1,{\cal{N}})&=&2{\cal{N}},
\nonumber\\
{\cal{Z}}(2,{\cal{N}})&=&{\cal{N}}(2{\cal{N}}-3),
\nonumber\\
{\cal{Z}}(3,{\cal{N}})&=&\frac{1}{3!}2{\cal{N}}(4{\cal{N}}^2-18{\cal{N}}+20),
\nonumber\\
{\cal{Z}}(4,{\cal{N}})&=&\frac{1}{4!}2{\cal{N}}(8{\cal{N}}^3-72{\cal{N}}^2+214{\cal{N}}-210),
\nonumber\\
{\cal{Z}}(5,{\cal{N}})&=&\frac{1}{5!}2{\cal{N}}(16{\cal{N}}^4-240{\cal{N}}^3+1340{\cal{N}}^2-3300{\cal{N}}+3024),
\label{a.07}
\\
{\cal{Z}}(6,{\cal{N}})&=&\frac{1}{6!}2{\cal{N}}(32{\cal{N}}^5-720{\cal{N}}^4+6440{\cal{N}}^3-28620{\cal{N}}^2
+63188{\cal{N}}-55440),
\nonumber\\
{\cal{Z}}(7,{\cal{N}})&=&\frac{1}{7!}2{\cal{N}}
(64{\cal{N}}^6-2016{\cal{N}}^5+26320{\cal{N}}^4-182280{\cal{N}}^3
+706216{\cal{N}}^2-1451184{\cal{N}}+1235520),
\nonumber\\
{\cal{Z}}(8,{\cal{N}})&=&\frac{1}{8!}2{\cal{N}}(128{\cal{N}}^7-5376{\cal{N}}^6+96320{\cal{N}}^5-954240{\cal{N}}^4
+5645192{\cal{N}}^3-19941264{\cal{N}}^2+38943000{\cal{N}}-32432400).
\nonumber 
\end{eqnarray}
The data shown in Fig.~\ref{fig003} are based on Eq.\ (\ref{a.07}).
\end{widetext}

Let us calculate the (not normalized) grand-canonical correlation function $\langle S_0^zS_j^z\rangle_z$,
where $S_j^z$ is the $z$-component spin operator of the trap $j$ and the subscript $z$ denotes the activity.
Defining a matrix
\begin{eqnarray}
{\bf{S}}=
\left(
\begin{array}{ccc}
0 & 0           & 0 \\
0 & \frac{1}{2} & 0 \\
0 & 0           & -\frac{1}{2}
\end{array}
\right)
\label{a.08}
\end{eqnarray}
one passes again from the sum over hard-dimer ground states for a fixed number of electrons $n$
and the sum over $n\le{\cal{N}}$
to the sum over
$s_0=0,\uparrow,\downarrow$,
$s_1=0,\uparrow,\downarrow$,
\ldots,
$s_{{\cal{N}}-1}=0,\uparrow,\downarrow$.
As a result
\begin{eqnarray}
\langle S_0^zS_j^z\rangle_z
={\rm{Tr}}\left({\bf{S}}{\bf{T}}^j{\bf{S}}{\bf{T}}^{{\cal{N}}-j}\right).
\label{a.09}
\end{eqnarray}
Using a MAPLE code
we can easily compute $\langle S_0^zS_j^z\rangle_z$ according to Eq.~(\ref{a.09})
for sufficiently large systems (up to ${\cal{N}}=256$);
the resulting expression for $\langle S_0^zS_j^z\rangle_z$ is a polynomial with the powers of $z$ from 2 to ${\cal{N}}$.

Turning to the canonical description we use the relation
\begin{eqnarray}
\langle S_0^zS_j^z\rangle_z
=\sum_{n=2}^{\cal{N}}z^n {\cal{Z}}(n,{\cal{N}}) \langle S_0^zS_j^z\rangle_n
\label{a.10}
\end{eqnarray}
[obviously
$\langle S_0^zS_j^z\rangle_{n=0}=\langle S_0^zS_j^z\rangle_{n=1}=0$
do not enter the right-hand side of Eq.\ (\ref{a.10})]
to derive
\begin{eqnarray}
{\cal{Z}}(n,{\cal{N}}) \langle S_0^zS_j^z\rangle_n
=\frac{1}{n!} \left.\frac{d^n\langle S_0^zS_j^z\rangle_z}{dz^n}\right\vert_{z=0}.
\label{a.11}
\end{eqnarray}
Thus the coefficients associated with the corresponding powers of the activity $z$ 
in the right-hand side in Eq. (\ref{a.09})
yield the required quantities $\langle S_0^zS_j^z\rangle_n$.

Finally,
we mention that the same results for $\langle \bm{S}^2\rangle_n$ can be obtained within the hard-dimer picture.
Since each hard-dimer ground state can be represented in terms of the occupation numbers of hard dimers on a simple chain,
i.e., it is enumerated by a set of hard-dimer occupation numbers $n_0,n_1,\ldots,n_{2{\cal{N}}-1}$, $n_j=0,1$,
and
$S^z=(1/2)\sum_{j=0}^{2{\cal{N}}-1}(-1)^jn_j$,
we have
\begin{eqnarray}
\langle {S^z}^2\rangle_n
=\frac{{\cal{N}}}{2}\sum_{q=0}^{2{\cal{N}}-1}(-1)^q\langle n_0n_q\rangle_n,
\label{a.12}
\end{eqnarray}
where $\langle \ldots \rangle_n$ in the right-hand side in Eq.\ (\ref{a.12})
stands for the (normalized) average over spatial configurations of $n$ hard dimers on a simple chain of $2{\cal{N}}$ sites.
To find a density-density correlation function at distance $q$
for one-dimensional hard dimers in the canonical ensemble $(n,2{\cal{N}})$,
it is convenient to calculate first a density-density correlation function at distance $q$ for one-dimensional hard dimers
in the grand-canonical ensemble $(z,2{\cal{N}})$,
\begin{eqnarray}
\langle n_0n_q\rangle_z={\rm{Tr}}\left({\bf{N}}{\bf{D}}^q{\bf{N}}{\bf{D}}^{2{\cal{N}}-q}\right),
\nonumber\\
{\bf{D}}
=
\left(
\begin{array}{cc}
1 & \sqrt{z} \\
\sqrt{z} & 0 \\
\end{array}
\right),
\;\;\;
{\bf{N}}
=
\left(
\begin{array}{cc}
0 & 0 \\
0 & 1 \\
\end{array}
\right),
\label{a.13}
\end{eqnarray}
which gives the required $\langle n_0n_q\rangle_n$ after inverting the relation
$\langle n_0n_q\rangle_z=\sum_{n=2}^{{\cal{N}}}z^n {\cal{Z}}(n,{\cal{N}}) \langle n_0n_q\rangle_n$.

\section{Linear independence of trapped electron states}
\label{b}

We wish to clarify
whether the set of ground states 
constructed in Secs. \ref{sec:IVA} and \ref{sec:IVB} 
for given $n\le n_{\max}$ is linearly independent.
An affirmative answer for hard-dimer states comes from Ref.\ \onlinecite{schmidt}.
For the sawtooth chain and the kagom\'{e} chains
the localized $n$-electron states ($n=1,\ldots,{\cal{N}}$) are linearly independent,
which is connected with the fact that for all three lattices there are sites which are unique to each cell
(isolated class in the nomenclature of Ref.\ \onlinecite{schmidt}).

We can use the same arguments for the set of single-electron states
which consists of localized states on a diamond/hexagon plus one two-leg state.
For this purpose the two-leg state (\ref{3.05})/(\ref{3.09}) in the set of states can be replaced
by the state localized along one (e.g., lower) leg only.
As a result we again are faced with the case when there are isolated sites
[the sites belonging to another (upper) leg]
that yields linear independence of the considered set of single-electron states
and thus of $n$-electron states ($n=1,\ldots,{\cal{N}}+1$).\cite{schmidt}


\begin{thebibliography}{999}
\bibitem{footnote_small}
Interestingly,
for finite (large) ${\cal{N}}$,
however,
$\xi_{+}$ is the largest one among $\xi_i$
(when $i=\pm,3$ or $i=\pm,3,4,5,6$)
only if $z$ does not exceed some (large) value $z^{\star}$.
Therefore,
for fixed $\mu_0-\mu>0$ if $T$ becomes smaller than some (small) $T^{\star}=(\mu_0-\mu)/\ln z^{\star}$
a crossover takes place
which may produce some tiny features
of finite(large)-${\cal{N}}$ low-temperature thermodynamic quantities.
\end{thebibliography}
\end{document}